%% file: graph_formats.tex
\documentclass[journal]{IEEEtran}
\pdfoutput=1

\usepackage[cmex10]{amsmath}
\usepackage{graphicx}
\usepackage[font=footnotesize]{subfig}
\usepackage{cite}
\usepackage{fixltx2e} 
\usepackage{color}
\usepackage{rotating}
\usepackage[colorlinks=true,linkcolor=blue,citecolor=blue,urlcolor=blue,pdfauthor={Matthew Roughan},pdftitle={Unravelling Graph Exchange File Formats}]{hyperref}
\usepackage[table]{xcolor}
\definecolor{palepink}{rgb}{1.0,0.9,0.9}

\usepackage[utf8]{inputenc}
\usepackage{amsfonts}
\usepackage{newunicodechar}
\newunicodechar{✓}{\checkmark}

\usepackage{fixltx2e} 

\newlength{\figurewidth}
\newlength{\captionspace}
\newlength{\figstarspace}
\newcommand{\ie}{{\em i.e., }}
\newcommand{\eg}{{\em e.g., }}

\newcommand{\anonymize}[1]{anonymous}

\def\equationautorefname~#1\null{%
  (#1)\null
}

\begin{document}

\title{Unravelling Graph-Exchange File Formats}
\author{Matthew Roughan and
  Jonathan Tuke
  \thanks{M. Roughan and J. Tuke are with the School of Mathematical
  Sciences at the University of Adelaide.}
}
\maketitle

\begin{abstract}
  A graph is used to represent data in which the relationships between
  the objects in the data are at least as important as the objects
  themselves. Over the last two decades nearly a hundred file formats
  have been proposed or used to provide portable access to such
  data. This paper seeks to review these formats, and provide some
  insight to both reduce the ongoing creation of unnecessary formats,
  and guide the development of new formats where needed. 
\end{abstract}


\section{Introduction}


\IEEEPARstart{E}{xchange} of data is a basic requirement of scientific
research. Accurate exchange requires portable file formats, where
portability means the ability to transfer (without extraordinary
efforts) the data both between computers (hardware and operating
system), and between software (different graph manipulation and
analysis packages).

A short search of the Internet revealed that there are well over 70
formats used for exchange of graph data: that is networks of vertices
(nodes, switches, routers, ...) connected by edges (links, arcs, ...).

It seems that every new tool for working with graphs derives its own
new graph format. There are reasons for this: new tools are often
aimed at providing a new capability. Sometimes this capability is not
supported by existing formats. And inventing your own new format isn't
hard.

More fundamentally, exchange of graph information just hasn't been
that important. Standardised formats for images (and other consumer
data) are crucial for the functioning of digital society. Standardised
graph formats affect a small community of researchers and tool
builders. This community is growing, however, and the need for
exchange of information is likewise growing, particularly where the
data represent some real measurements that were expensive to collect.

The tendency to create new formats in preference to using existing
tools is unhelpful though, particularly as the time to ``create'' a
format might be small, but the time to carefully test formats and
read/write implementations is extensive.  Reliable code is critical to
maintain data quality, but many tool developers seem to focus on
features instead of well-audited code. Moreover support of formats,
for instance clear documentation and ongoing bug fixes, is often
lacking.

An explosion of formats is therefore a poor state of
affairs. The existing formats do include many of the features one might need,
and some are quite extensible, so the bottleneck is not the existing
formats so much as information about those formats.  This is the gap
this paper aims to fill.

This work concentrates on graph {\em exchange} formats. Such formats
have certain requirements above and beyond simple storage: most
obviously portability. However, portability in this context is not
purely about syntax. Exchange also requires common definitions of the
meaning of the attributes.



On the other hand, file size is not a primary consideration. Hence many
exchange formats pay little attention to this and related details
(e.g., read/write performance).  

We concentrate on exchange formats, but some of the formats considered
here were not originally developed with exchange in mind, but have
become {\em de facto} exchange formats through use. In these cases we
see reversals of objectives compared to some purpose-built exchange
formats.  We shall therefore consider a large range of such features
for comparison, noting as we do so that as exchange of very large
datasets becomes important, the requirements will change.

Many of the formats presented may seem obsolete. Some are quite old
(in computer science years). Some have clearly not survived beyond the
needs of the authors' own pet project. However, we have listed as many
as we could properly document, partially for historical reference, and
partially to show the degree of reinvention in this area. But more
importantly, because old and obscure isn't bad. For instance NetML, a
format that doesn't seem to be used at all by any current toolkits,
incorporates some of the most advanced ideas of any format
presented. A good deal could be learnt by current tool builders if
they were to reread the old documentation on this format.

It is important to note that this paper does not present yet-another
format of our own. It is common in this and other domains for the
discussion of previous works to be coloured by the need to justify the
authors' own proposals. Here we aim to be unbiased by the need to
motivate our own toolkit, and so (despite temptation) do not provide
any such.

We do not argue that new graph formats should never be
developed. In some applications new features are needed that are not
present in the existing formats. However, it is critical that those
who wish to propose new ideas should understand whether they are {\em
  really} needed. Moreover, in studying the existing formats, and their features,
we learn what should be required in any new format to make it more
than a one-shot, aimed at only one application. In fact, the results
suggest that new formats are desirable for several reasons, but that
perhaps what would be more useful would be a container format capable
of providing self-documentation and meta-data-like features, while
encapsulating a set of formats with variable levels of feature
support. 

So the value of this work is threefold: firstly it provides a
relatively complete set of information about the currently available
formats, secondly it provides a basis for selection of a suitable
format, and thirdly it provides information about the nature of the
features that can and have been used in future developments of graph
exchange strategies.

\section{Background}

Graphs (alternatively called networks) have been used for many years
to represent relationships between objects or people. 

A mathematical {\em graph} ${\cal G}$ is a set of nodes (or vertices)
${\cal N}$ and edges (or links or arcs) ${\cal E} \subset {\cal N}
\times {\cal N}$. 

An alternative representation of a graph can be given through its
{\em adjacency matrix} $A$, defined by
\[ A_{ij} = \left\{ \begin{array}{ll}
        1, & \mbox{ if } (i,j) \in {\cal E}, \\
        0, & \mbox{ otherwise. } \\
      \end{array} \right. 
\]
Other representations exist (and are discussed below in detail). These
alternatives are often used to create computationally efficient
operations on the graph. Underlying these alternatives is the choice
of the first-class objects to be represented: mathematically, the
graph is the first-class object, and nodes and edges are components of
the graph, but it is useful, for instance, to represent the edges as a
set of objects each with their own components (including their
end-points), or to represent the nodes as the first-class objects,
with edges as properties of the nodes. Each alternative has advantages
in terms of particular algorithms that can be applied.

Additional information is often added to a graph: for instance
\begin{itemize}

\item node or link labels (names, types, ...);

\item values (distances, capacity, size, ...); and

\item routing (paths taken when traversing the graph).

\end{itemize}
This additional information is often critical to make use of the graph
data in any real application.  

It has been necessary for many years for researchers in sociology,
biology, chemistry, computer science, mathematics, statistics and
other areas to be able to store graphs representing concepts as
diverse as state-transition diagrams, computer-software structure,
social networks, biochemical interactions, neural networks, Bayesian
inference networks, genealogies, computer networks, and many more.
Researchers also need to share data. They have done so by sharing
files. As a result portable file formats for describing graphs have
been around for decades.

This document is concerned with providing information about these
formats, specifically with the intention of moving towards a smaller
number of standard formats (the current trend seems to be progressing
in the other direction).

We only look here at publicly disclosed formats, for the obvious
reason that a format can't really be called a data exchange format
unless its definition is public. It is fair to say that although many
were intended for exchange of information, most failed at this and
were only really used for a single tool or database of graphs. In a
few other cases, the format was not intended as an exchange format,
but has become a {\em de facto} exchange format by virtue of the
inclusion of IO routines in other software than its originator. In any
case, we have tried to be inclusive here: we include anything that
might be reasonably called an exchange format (and which is publicly
documented to some degree), rather than trying to exclude those which
we guess are not.

There are many subtypes of graphs, and generalisations. For instance:
the general description above is that of a {\em directed} graph.  An
{\em undirected graph} has the property that if $(i,j) \in {\cal E}$
then so too is $(j,i)$.

It is important to note that it is often possible to represent one
type of graph in terms of the other: for instance an undirected graph
may be represented by a directed graph by including all reverse links
in the data. However, this is inefficient.
 
Moreover there is the issue of {\em intention}. The intention of the
person storing the data is important: for instance, an undirected
graph that is stored as a directed graph may be edited to become
directed. A native undirected format enforces the correct
semantics. Thus when considering the type of graph being stored, we
consider the native or explicitly supported subtypes, not those that
can be implicitly supported. 

Other generalisations of graphs include multi-graphs, hyper-graphs,
and meta-graphs (described in more detail below). Subtypes include
trees and DAGs (Directed Acyclic Graphs). Once again, it is often
possible to represent these in terms of the simple directed graph, but
often this will be inefficient, and deficient in terms of intention. We
will therefore look for native support for these generalisations and
subtypes.

\subsection{Related work}

We distinguish this work from the study of {\em graph databases},
which have a similar role in storing data where the relationships have
at least as much importance as the entities they relate. However,
although they may hold the same type of data that we are considering
here, the motivations for a graph database are different. Typically,
those concerned with databases are interested in ACID (Atomicity,
Consistency, Isolation, Durability) and other similar properties. The
underlying assumption is that the data is changing dynamically
according to some set of transactions and operations and that the
database should work correctly under these conditions. Consequently
graph databases are not simply concerned with the structure and
description of the data, but also how that data may be operated on,
and queried.  On the other hand, the standard assumption in data
exchange is that the data itself is relatively static, but portability
is important.

There is a wide-ranging survey of graph databases~\cite{Angles:2008},
which is more concerned with the underlying database aspects, \eg the
relationship between a graph database and other more traditional
databases such as a relational database, and the properties of various
exemplar graph databases. 

There is some overlap of concerns: in both cases there is some
interest in data integrity, compression, and the like, but it is fair
to say that these issues have typically taken second place in the
design of graph exchange formats. 

There have been a number of other efforts to gather similar
information on graph exchange formats by researchers
\cite{bodlaj13:_networ,bernard:_graph_file_format,10:_netwik} and
software distributors \cite{yfiles_io,gephi}. The results provided
inspiration for some of the descriptors used here, but this paper aims
to provide a more comprehensive summary.

One additional paper to consider is
\cite{brandes00:_graph_data_format_works_repor}, which was written
specifically with the view of designing a new, more universal graph
format. We deliberately avoid this approach in order to avoid bias in
our discussion.

\clearpage
\section{The File Formats}

As noted the aim here is to describe graph {\em exchange} formats, \ie
formats that are used to exchange data between scientists and
programming environments. Not all of the formats started out that way
-- some were intended as internal formats for a particular software
system, but have become {\em de facto} exchange formats when another
system sought to leverage existing data by incorporating an existing
format. A few of the formats are still primarily internal to a single
system, but are important to describe because they exhibit an
interesting feature. In the main we concentrate on those that were
designed with data exchange in mind, or have been used in that way in
practice.

This list is incomplete. There are some formats that we have observed
in the literature, but have been unable to find documented (\eg
Gem2Ddraw), or which appear to only be used as an internal format for
a single tool.  The graph formats we know of that have been excluded
are the Tom Sawyer format, gem3Ddraw, PROGRES, GTXL, GedML, UXF, GRL,
VEGA, BLGF, GraphLab, BNIF, BIF, XGML, NMF, Inflow, GDS, Tnet and
RDF. Additional information sources covering these would be welcomed.

There are a few formats that we have lumped together under the general
heading of {\em TGF} (the Trivial Graph Format) because they are all
functionally equivalent to a delimited edge list. There is no point
in listing every variant of this approach: there are many and they
vary mainly on the choice of storage (plain ASCII through to Excel),
and delimiter (tabs and commas are common).

There are many file formats that could, in principle, contain a graph:
\eg XML, JSON, SGML, Avro, YAML (YAML Ain't Markup Language), RDF (the
Resource Description Framework), HDF (the Hierarchical Data
Format). For that matter any image file could contain an adjacency
matrix. Unless there is a specific extension of these designed to
provide support for graphs, in which case we list the specific not the
generic. For instance, several software tools say that they can
read/write JSON or other generic serialisations of data, but without
details of exactly what is being serialised, then these are not useful
exchange formats. We treat Matlab's {\tt .mat} format as a special
case because it has explicitly been used to exchange graph data, at
least between instances of Matlab, even though it is a generic data
format.

We also aim to avoid, for simple practicality, formats that represent
data that has a graph structure, but whose main content is not the
graph. For instance HTML: the graph structure of the WWW is vastly
smaller than the content and HTML is intended to store both in a
distributed fashion. If one wished to represent the graph of the WWW,
then another format seems indicated. Other examples include SBML (the
Systems Biology Markup Language), and FOAF \cite{foaf} (Friend Of A
Friend).  

\autoref{tab:formats} provides the list of exchange formats we do
include, as well as links and references. Check marks in the table
indicate that we have had at least cursory feedback about the
information in the table from one of the creators or maintainers of
the format (we received such feedback on 23 of the formats). Please
see the acknowledgements for a list of contributors.
 
We have also tried to include a {\em reference time frame} to provide
some historical context for the format. The dates are based on
explicit records from the first recorded reference to the format,
through to the last recorded date of maintenance. However, this
information is often not supplied, so we have used the closest
available proxy. For instance, change-logs or copyright dates on
format documentation or publication dates for papers.  Hence these
should not be seen as a completely reliable data. It is an attempt to
document the historical development of this field, so much of which is
not in the archival journals\footnote{Please note that some formats
  that are notionally obsolete according to reference dates, but may
  still be used by archival stores of graph data.}.

\begin{figure}[htbp]
	\centering
	\includegraphics[width=0.95\columnwidth]{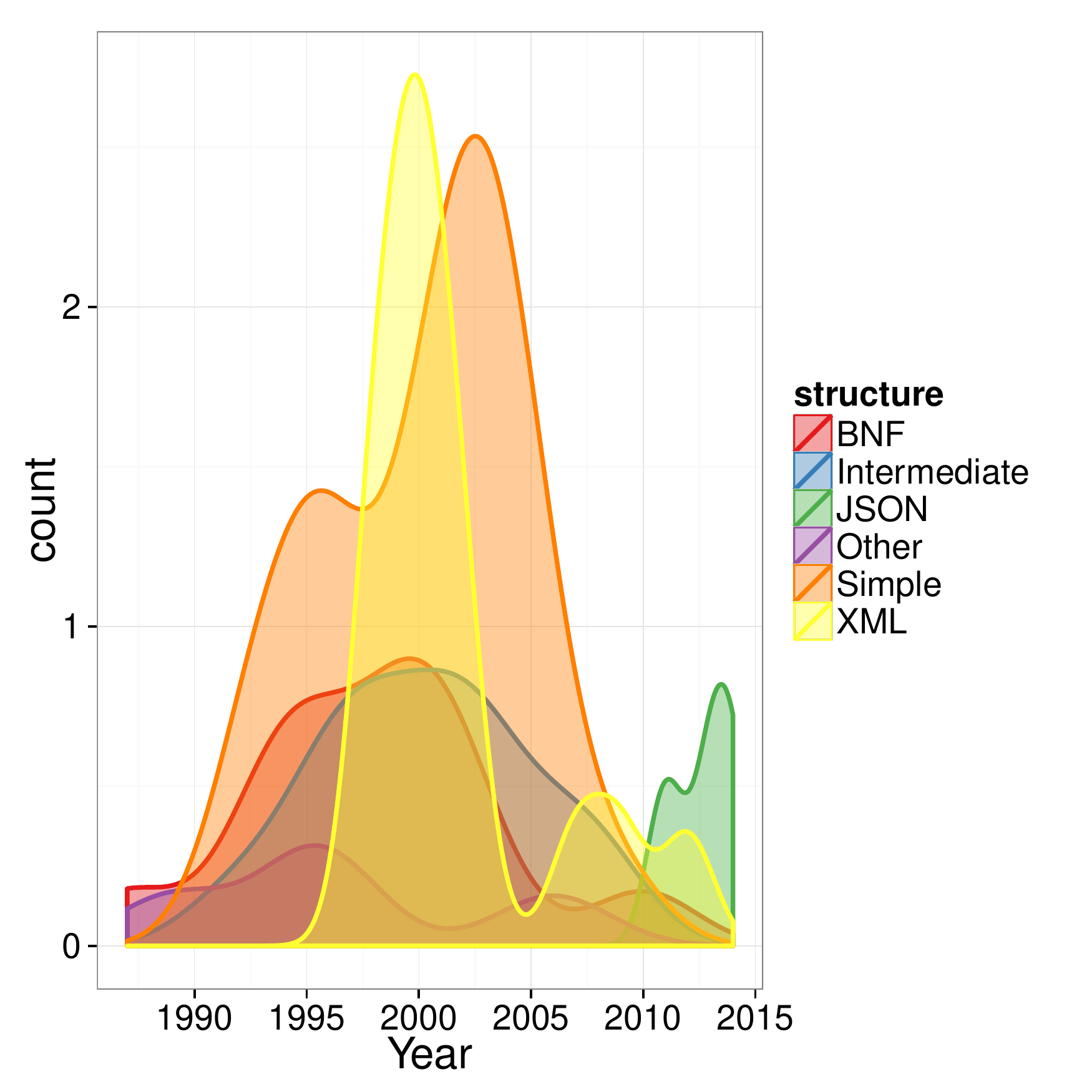}
	\caption{New format origination dates.} 
	\label{fig:time_plots}
\end{figure}

For instance, \autoref{fig:time_plots} provides a quick summary. We
can see that there was a flurry of activity in the late 90s continuing
on until today, but the style of contributions has changed over
time. It is interesting to see how XML became flavour of the day
around in the late 90s, and then dropped out of popularity in recent
years, and in the most recent past there seem to be several efforts to
design graph formats on top of JSON. It seems there are fads even
within technical fields.


\rowcolors{2}{white}{palepink}

\begin{table*}[p]
  \hyphenpenalty=100000
  \centering
  {\footnotesize
    \begin{tabular}{rrrllr@{ -- }l}
      \input{excel2latex/tab0.tex}
      \hline
      \input{excel2latex/tab1.tex} 
    \end{tabular} 
  } 
  \caption{The format list. Checkmarks indicate  formats that have had
    their details audited by someone associated with creation or
    maintenance of the format.}
  \label{tab:formats}
\end{table*}
 
\clearpage
\section{Descriptors and Discriminators}

In order to describe the formats we will consider here, we need some
simple means to compare and contrast.  Of a necessity, these will
oversimplify some of the issues. For instance, where a format uses
multiple files we have not attempted to explain exactly how data is
divided between these files. 

What's more, many descriptions of file formats are imprecise. It is
common to describe the format by reference to examples. Although
useful for simple cases, these leave out important details: for
instance: the character set supported, and even more surprisingly, the
format of identifiers. It is often vaguely suggested that these are
numbers, but without formal definition of what is allowed (presumably
non-negative integers, but are numbers outside the 32 bit range
supported?). 

In the following, we make the best estimate of the capabilities of
each format through reference to online documentation, and through a
survey of the file format creators. In many cases the results are
inferences, so in this section we will outline the features we
describe, and the assumptions made in compiling our data. However, we
have made the best effort possible to contact authors of formats, and
their comments about capabilities have been given precedence.

There are three main types of descriptors here: 
\begin{LaTeXdescription}

\item[file type]: these are simple issues of the type of file storing
  the data: binary vs ASCII, etc.
 
\item[graph types]: this refers to the nature of the graph data that
  can be stored.

\item[attributes]: these are features related to supplemental data
  about nodes and edges, such as labels and values associated with
  these.

\item[general]: this is a grab bag for additional features that don't
  fit in either of the previous classes.

\end{LaTeXdescription}
We'll describe each of these in detail below, and then provide a table
of the features vs file formats.

One last point, this is not intended as a pejorative list. We do not
mean to imply that having a feature is good or bad. The aim is to
provide potential users with the background to choose the right format
for their purposes. 

\subsection{File Type}
\label{sec:file_types}

\begin{LaTeXdescription}
\item[encoding]: This is, in principle, a simple distinction in file
  type between text and binary files. However, text files today can
  use multiple different character sets, and this is important because
  some graphs will be labelled with non-English character
  sets. However, the majority of file-format definitions leave
  unspecified the character set to be used. We assume here that the
  character set is ASCII, unless there is some indication otherwise,
  either an explicit statement, or in the case of applications of XML
  it is assumed that the character set supported is
  Unicode. \autoref{fig:char_supp_prop} indicates the proportions of
  files providing each type of encoding.

\begin{figure}[htbp]
  \centering
  \vspace{-10mm}
  \includegraphics[width=0.95\columnwidth]{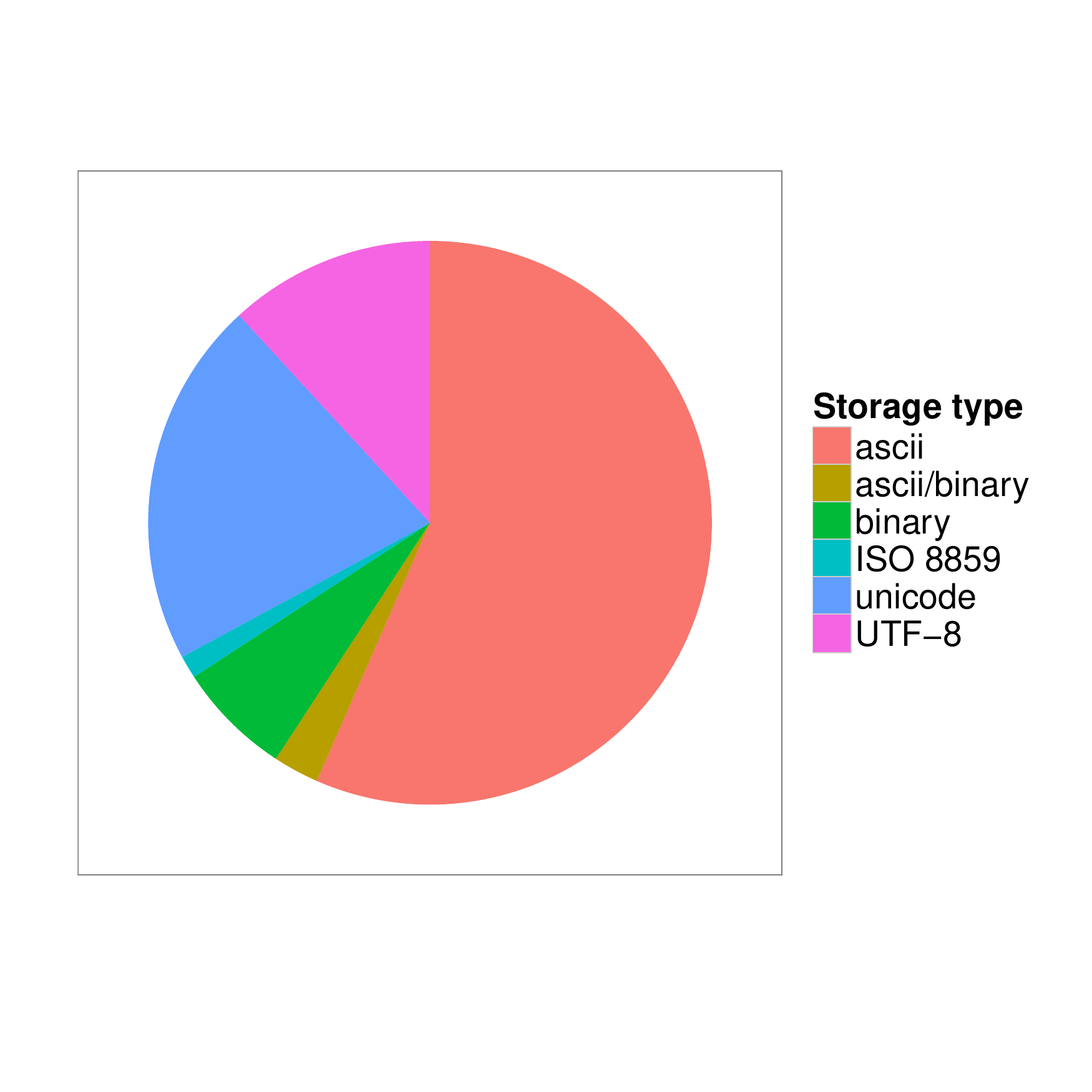}
  \vspace{-16mm}
  \caption{Support for different encodings.}
  \label{fig:char_supp_prop}
\end{figure}

\begin{figure*}[t]
  \centering
  \rowcolors{2}{white}{white}
  \begin{minipage}[t]{0.3\linewidth}
    \includegraphics[width=\textwidth]{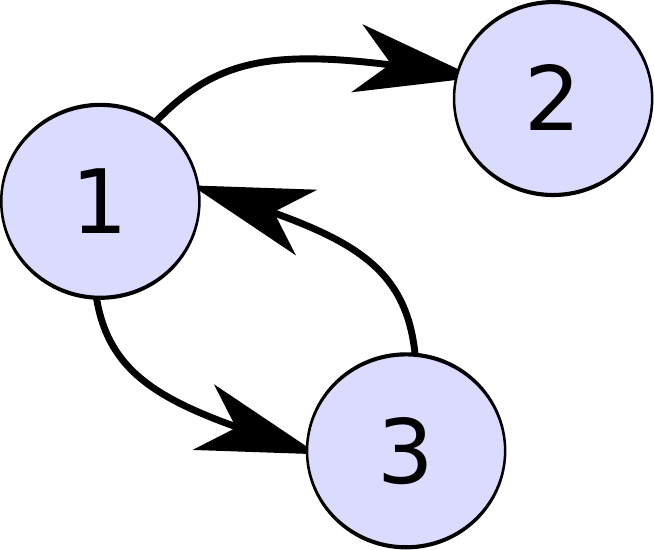}
  \end{minipage}
  \\[3mm]

  \newlength{\currentfiglenth}
  \setlength{\currentfiglenth}{0.24\linewidth}

  \begin{minipage}[b]{\currentfiglenth}
    \centering
    \[ A = \left( \begin{array}{rrr}
        0 & 1 & 1 \\
        0 & 0 & 0 \\
        1 & 0 & 0 \\ 
      \end{array} \right) \]
    (a) Adjacency matrix
  \end{minipage}
  \begin{minipage}[b]{\currentfiglenth}
    \centering
     \[
      \begin{array}{ll}
        1,2 \\
        1,3 \\
        3,1 \\
      \end{array}
      \]

     (b) Edge list.
  \end{minipage}
  \begin{minipage}[b]{\currentfiglenth}
    \centering
     \[
      \begin{array}{ll}
        1: & 2, 3 \\
        2: &  \\
        3: & 1 
      \end{array}
      \]

      (c) Neighbour lists.
  \end{minipage}
  \begin{minipage}[b]{\currentfiglenth}
    \centering
     \[
      \begin{array}{ll}
        3,1,2 \\
        1,3 \\
      \end{array}
      \]

      (d) Paths.
  \end{minipage}

  \caption{Simple directed graph with representations.}
  \label{fig:directed_graph}
\end{figure*}

\item[representation]: Methods to represent a graph include:
  \begin{LaTeXdescription}
  \item[matrix]: The graph's full adjacency matrix.

  \item[edge]: A list of the graph's edges~\cite{ebert87:_versat_data}.

  \item[smatrix]: The matrix representation is poor for sparse graphs,
    which are common in real situations. However, some tools actually
    store a sparse matrix, which is almost equivalent to an edge
    list\footnote{There is one exception to this: Cluto stores sparse
      matrices in a format more closely resembling the neighbour
      representation.}. There is a subtle difference in that a matrix
    view of the edges in a network cannot contain much detail about
    the edges (only one number), and so we have a separate name, {\em
      smatrix}, for formats that use this type of representation.

  \item[neighbour lists]: This is a list of the graph's nodes, each
    giving a list of neighbours for each node. Often called {\em
      adjacency lists} we avoid that term because it is easily
    confused with the edge list.

  \item[path]: One can also implicitly represent a graph as a series
    of {\em path} descriptions (essentially a path is a list of
    consecutive edges). This could be useful, for instance, with a
    tree or ring. 

    Moreover, graph data is often derived from path data, \ie a series
    of paths are analysed, and the edges on these become the graph. In
    other cases, one might like to store path information, for
    instance related to routes along with the graph.

  \item[constructive]: Graphs can often be described in terms of
    mathematical operations used to construct the graphs: for instance
    graph products on smaller graphs
    \cite{parsonage11:_gener_graph_produc_networ_desig_analy}. See
    \cite{batagelj95:_towar_netml} for a description of ``levels'' of
    graph formats.

    Apart from simple incremental construction, the only format that
    seems to allow this is NetML~\cite{batagelj95:_towar_netml}.

  \item[procedural]: Many graphs can be concisely defined by a
    set of procedures, rather than explicit definition of the nodes
    and links. This type of graph format could be very concise, but
    verges on creating another programming language. In fact, many
    graph libraries for particular programming languages essentially
    provide this, but in a non-portable manner. 

    The only {\em generic} (language independent) format that seems to
    allow this is NetML~\cite{batagelj95:_towar_netml}.

    Any procedural approach admits the possibility of defining a
    method for {\em constructive} graph description, but we do not
    automatically count any procedural approach as constructive,
    unless it provides explicit graph-related operations as part of
    the toolkit.


  \end{LaTeXdescription}
  These representations are given varying names in the literature, but
  we use the names above to be clear. \autoref{fig:directed_graph}
  illustrates four of these, and \autoref{fig:representation_prop}
  shows the proportions 

  \begin{figure}[t]
    \centering
    \vspace{-10mm}
    \includegraphics[width=0.95\columnwidth]{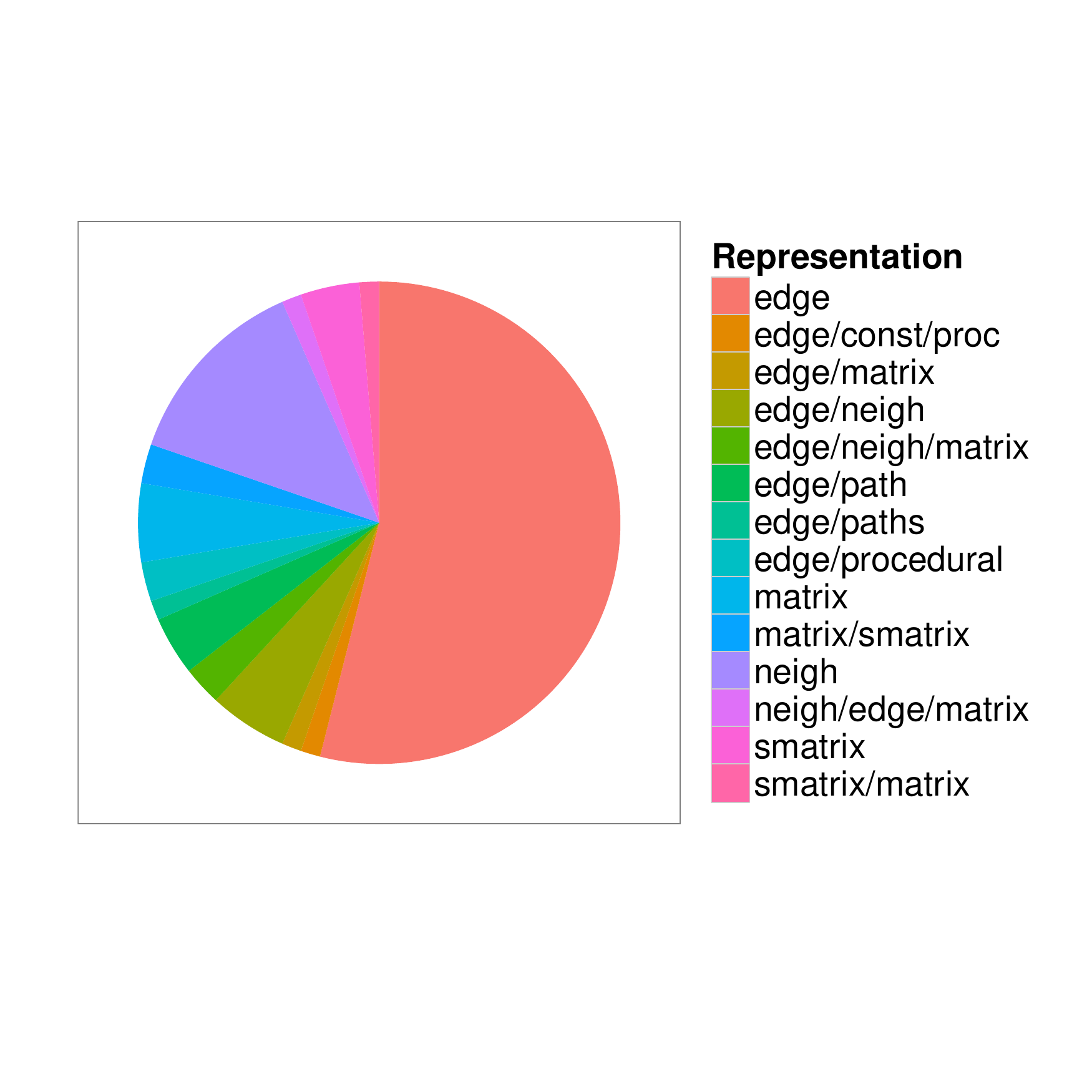}
    \vspace{-16mm} 
    \caption{Proportions supporting different representations.}
    \label{fig:representation_prop}
  \end{figure}

  The representation is important: for a graph with $N$ vertices and
  $E$ edges, the adjacency matrix requires $O(N^2)$ terms, the edge
  list $O(E)$ terms, and the neighbour list $O(N + E)$ terms. However,
  the terms in a matrix are $\{0,1\}$ whereas the terms in the edge
  and neighbour lists are node identifiers (consider they might be 64
  bit integers), so the size of a resulting file based on each
  representation depends on many issues, including the way the data is
  stored in the file. No approach is universally superior.

  Moreover, some may be easier to read and write: for instance a
  neighbour listing may be slightly more compact than an edge list,
  but the latter has the same number of elements per line, potentially
  making it easier to perform IO in some languages.

  More subtly, a neighbour-list representation treats edges as
  properties of nodes, whereas an edge list treats edges as objects in
  their own right; and the matrix representation treats the graph as
  the only object with nodes and edges as properties of the graph.
  Although a program can internally represent data however it likes,
  and read in a neighbour list into structures that treat edges as
  objects in their own right, the native treatment of data is
  reflected in the ease with which attributes can be added. For
  instance, in a neighbour list it is intrinsically harder to record
  attributes for edges, and in the matrix representation it is harder
  to record attributes for nodes. This is, fundamentally, why we regard
  edge-list and sparse-matrix formats as different.
 
  Some graph file formats allow alternative representations, and so we
  list all that are possible. However note that this is often actually
  multiple file formats under one name. It seems rare to allow a mixed
  representation.

  We haven't (yet?) reported on whether edge-list formats explicitly
  lists nodes or only implicitly lists them as a consequence of
  edges. The latter is briefer, but requires a special case for degree
  0 nodes.
 
  When considering generalisations of graphs, other representations
  are possible (for instance tensors can generalise the concept of an
  adjacency matrix for multi-layer networks). However, codification of
  these is an ongoing research topic \cite{kivela:_multil} and so we
  will not try to encapsulate it here.

\item[structure:] This field describes how the file format's structure
  is defined. The cases are:
  \begin{LaTeXdescription}

  \item[simple]: the typical approach to create a graph format is to
    use one line per data item (a node, an edge, or a neighbourhood),
    with the components of a line separated by a standard delineator
    (a comma, tab, or whitespace). There are many variations on this
    theme, some more complex than others, for instance including
    labels, comments or other information. These formats are usually
    specified by a very brief description and one or two
    examples. They rarely specify details such as integer range or
    character set.

  \item[intermediate]: this is a slight advance on a {\em simple} file
    format, in that it includes some grammatical elements. For
    instance, the file may allow definition of new types of labels for
    objects. However, in common with simple files, these are usually
    only specified by a very brief description and one or two
    examples, not a complete grammar.

  \item[BNF]: means that the file format is described using a grammar,
    loosely equivalent to a Backus-Naur Form (BNF). This is perhaps
    the most concise, precise description. When done properly it
    precisely spells out the details of the file in a relatively short
    form.


  \item[XML, JSON, SGML, ...]: many graph file formats extend XML,
    JSON, SGML, or similar generic, extensible file formats. This is a
    natural approach to the problem, and allows a specification as
    precise as BNF, though only through reference to the format being
    extended. Thus it is precise, but sometimes rather difficult to
    ascertain all of the details, unless one is an expert in XML, etc.

    On the other hand, these approaches draw on the wealth of tools
    and knowledge about these data formats. On the other hand again,
    to use those tools the model of your graph object has to map to
    the XML model (or at least be easily transformed into that form).

  \item[Tcl, Lisp, ...]: As noted above one approach to defining a
    graph is procedural. Most of the approaches that allow this are
    extensions or libraries for common programming languages. 

    We will not list every programming language and library as a data
    format though because, generically, such approaches are not
    portable between programming languages. We do mention a few
    formats though (ns-2 and S-Dot), because translators exist
    from/or to these from other data formats.

  \end{LaTeXdescription}
  \autoref{fig:structure_prop} shows the proportion of each type of
  structure within the files.
  
  \begin{figure}[t]
    \centering
    \vspace{-10mm}
    \includegraphics[width=0.95\columnwidth]{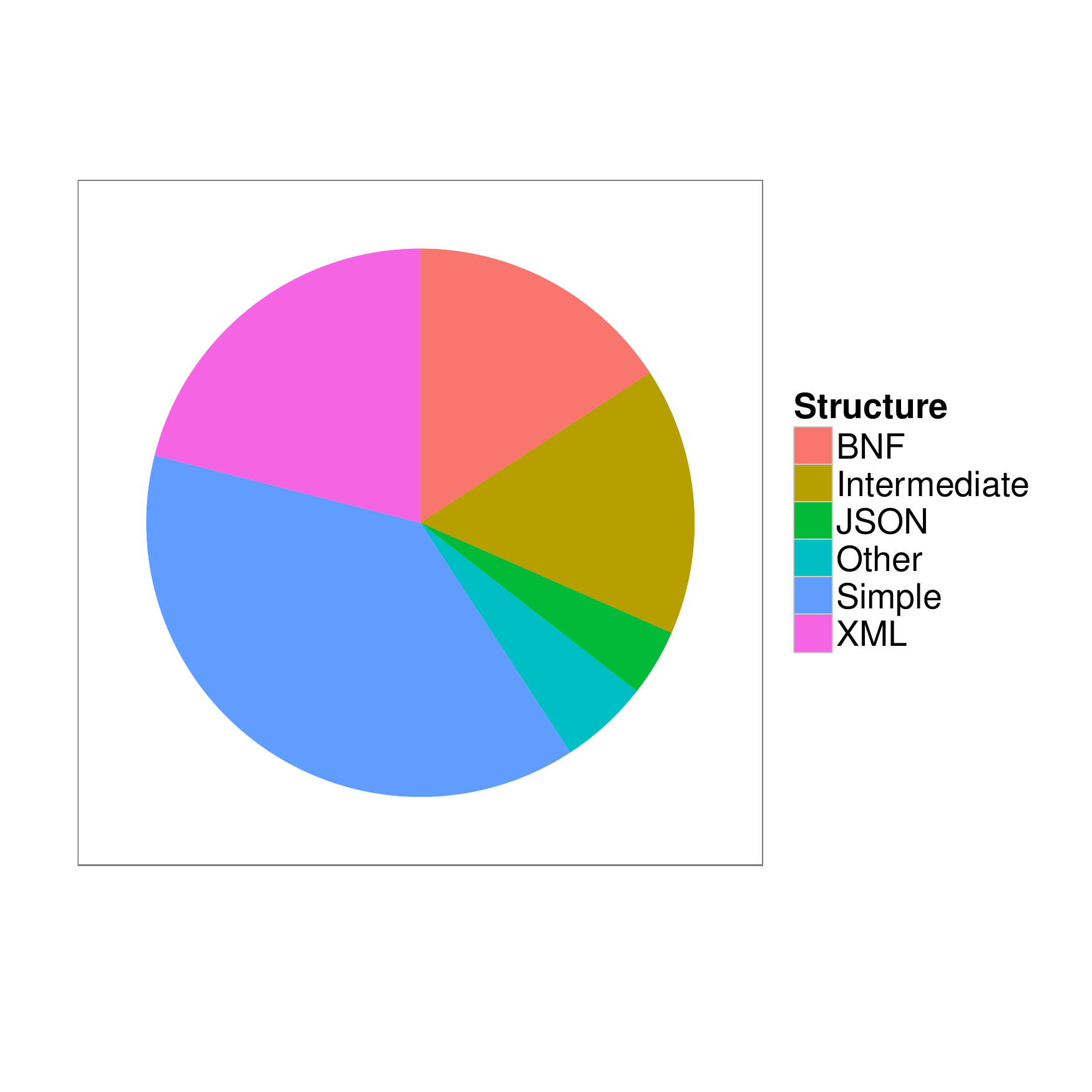}
    \vspace{-16mm}
    \caption{Proportions with each structure.}
    \label{fig:structure_prop}
  \end{figure}
  
\item[single or multiple files]: Most file formats use a single file,
  but some formats require multiple, for instance, a separate files
  for the lists of nodes and edges. Other formats allow supplementary
  information in additional files, so multiple files aren't
  mandatory. We have only classified the files by whether multiple
  files are allowed, not whether they are mandatory (because the
  latter requires a distinction about what mandatory would mean: does
  it mean they are required to support basic features or advanced
  features?)



\item[integral meta-data]: Meta-data is data about the graph: for
  example, its name, its author, the date created, and so on. This is
  very important data, but many formats provide no means to include it
  in the file, and instead rely on external records. We refer to
  meta-data as {\em integral} if it is contained in the file itself.  

  Some formats allow meta-data through unstructured comments. This is
  better than nothing, but lack of structure of the comments means
  these are not machine readable, in general.

  Some file formats provide only a limited range of meta-data fields,
  whereas others are arbitrarily extensible. To distinguish the
  various cases we fill this field with one of the following:
  \begin{LaTeXdescription}
  \item[no]: No meta-data is allowed.
  \item[comments]: Unstructured meta-data is allowed in comments.
  \item[fixed]: A defined set of meta-data can be included, \eg a
    date or name field is predefined as part of the format.
  \item[arbitrary]: An explicit mechanism is described to allow
    the user to specify arbitrary meta-data to be included.
  \end{LaTeXdescription}
 
  The value of meta-data is clear, but once again, let us reiterate
  that there are plusses and minuses in different approaches.  For
  instance, arbitrary meta-data may seem superior, but can then lead
  to ambiguity about what meta-data should be kept for each dataset,
  whereas having a fixed list of attributes can make it obvious what
  is expected. However, it is common for formats to have support
  downwards, \eg formats with fixed attributes often also support
  comments, and those with arbitrary properties often support some set
  of fixed properties and comments.

  \autoref{fig:meta-data_prop} shows support for various types of
  meta-data in the formats. 

\begin{figure}[htbp]
  \centering
  \vspace{-10mm}
  \includegraphics[width=0.95\columnwidth]{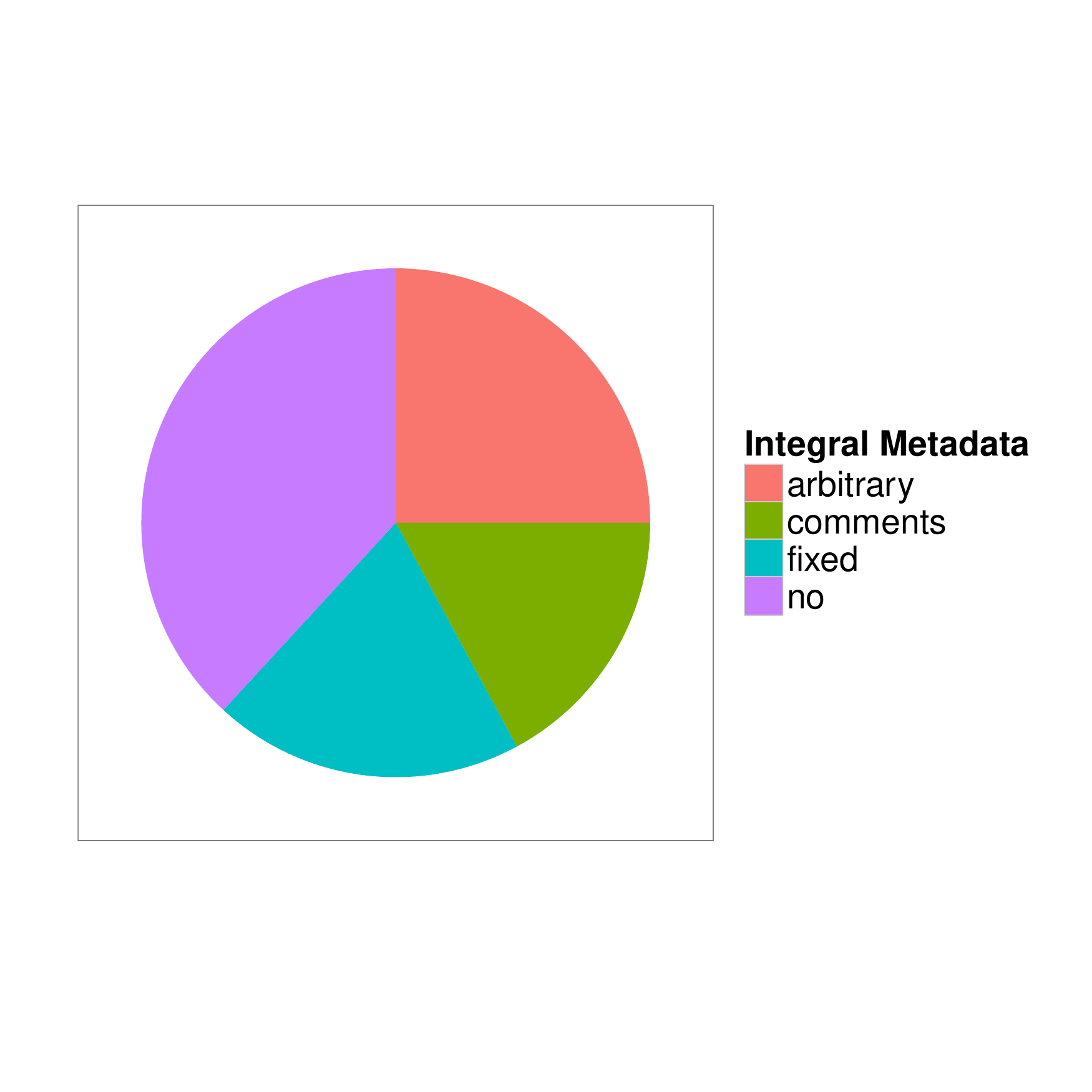}
  \vspace{-16mm}
  \caption{Support for meta-data.}
  \label{fig:meta-data_prop}
\end{figure}

\item[built-in compression]: It is easy enough to compress a
  graph-file using common utilities such as gzip, and typical
  compression ratio will be reasonably good as graph files often have
  many repeated strings. However, one format (BVGraph) provides for
  compression of the graph as it is written, in much the way image
  file formats allow intrinsic compression of the image. 

  Graph Compression algorithms have been a topic of study at least
  since 2001
  \cite{Adler:2001:TCW:882454.875027,Randall:2002:LDF:882455.874988,boldi04:_bvgraph},
  with numerous followups. So it is interesting that only one format
  is designed around this feature. However, two other formats provided
  some crude mechanisms to reduce the size of the file. Finally DGS
  formally acknowledges the role of compression by requiring that a
  gzipped file be accepted by software reading its format.


\end{LaTeXdescription}
\autoref{tab:file_types} provides the information on file types.

\begin{table*}[p]
  \hyphenpenalty=100000
  \centering
  {\footnotesize
    \begin{tabular}{r|llllp{15mm}p{17mm}}
      \input{excel2latex/tab2.tex}
      \hline
      \input{excel2latex/tab3.tex} 
    \end{tabular} 
  }
  \caption{File types (see \autoref{sec:file_types} for explanation of columns).}
  \label{tab:file_types}
\end{table*}

\subsection{Graph Types}
\label{sec:graph_types}

\begin{LaTeXdescription}
\item[directed/undirected]: The two basic forms of graph are the
  directed and undirected graph. In the former edges (or arcs) imply a
  relation from one node to another. In the later an edge implies a
  relationship in both directions.

  Some graph formats specify one or the other; others allow the user
  to specify either, and the most general allow the user to specify
  the type of each edge\footnote{Of course a directed graph format can
    contain an undirected graph by including edges in both directions,
    but we are considering here whether it can do this a little more
    succinctly.}. In one case, the format is explicitly restricted to
  DAGs (Directed Acyclic Graphs).

  Many graph formats fail to specify their type. In that case we
  assume it is {\em directed} if the edges/arcs are specified by
  directional nomenclature (\eg from/to or source/destination). We
  also assume that matrix formats are directed unless there is
  specific mention of mechanism to represent the upper triangular part
  of the matrix alone.

\begin{figure}[htbp] 
  \centering
  \vspace{-10mm}
  \includegraphics[width=0.95\columnwidth]{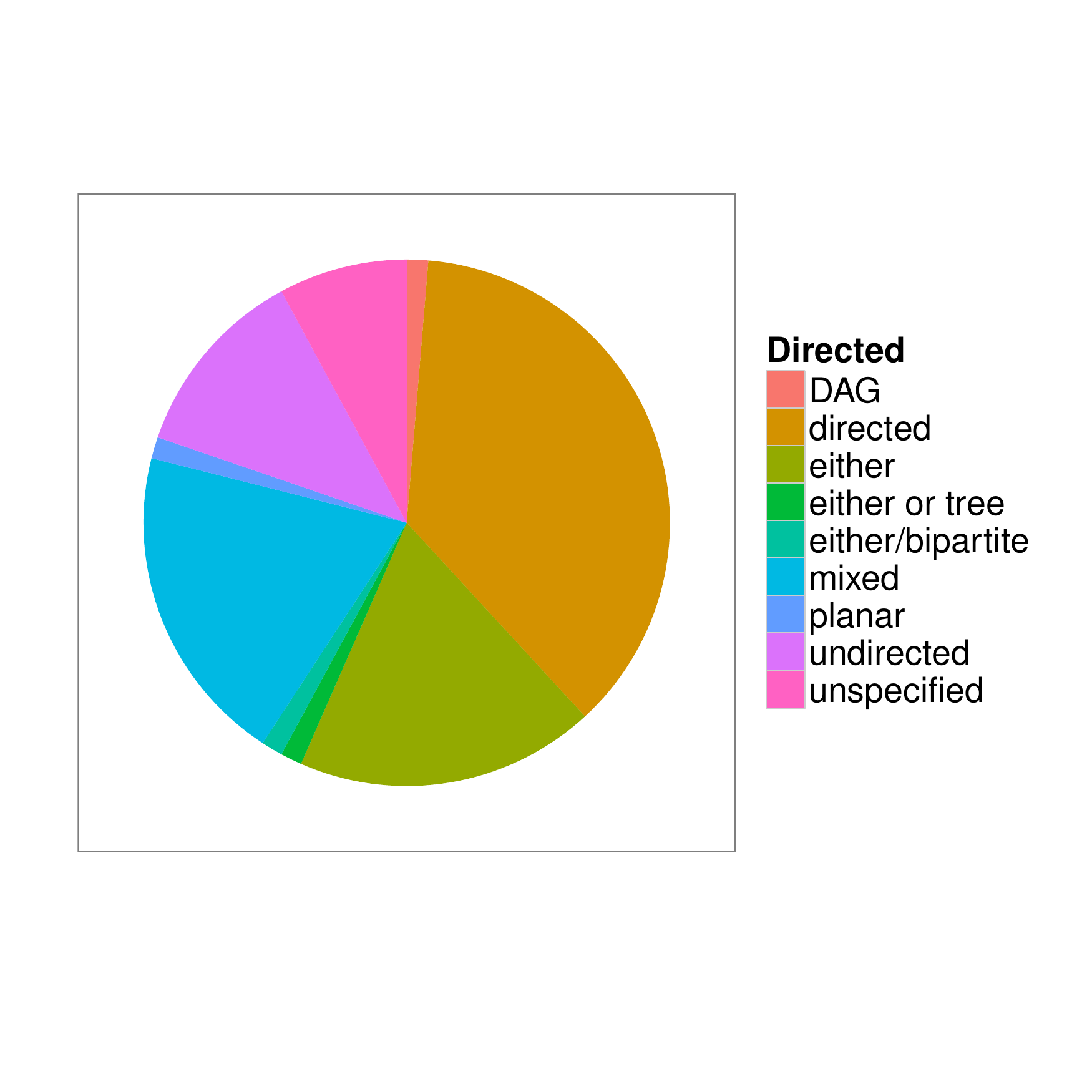}
  \vspace{-16mm}
  \caption{Graph type support.}
  \label{fig:multi_attr_prop}
\end{figure}

\item[multi-graph]: A multi-graph is a graph generalisation that
  allows (i) self-loops, and (ii) more than one edge between a single
  pair of nodes.

  Some formats specifically allow, or disallow multi-graphs. A few
  allow loops, but not multi-edges. Many, however, say nothing on the
  topic. We assume in this case that formats presenting either matrix
  or neighbour lists representations don't allow multi-graphs. It is
  technically possible to represent a multi-graph in these cases, but
  this would require special processing of the information, and unless
  we see an indication this is present we assume it is not. Edge
  lists, however, can easily cope with multi-graphs. We suspect it is
  left to the software supporting the data format to make a decision
  about how to deal with these cases, and the decision may be
  inconsistent between supporting software. Hence it seems important
  that when an edge-based representation leaves the question
  unspecified, we note this status.

\item[hyper-graphs]: A hyper-graph allows edges that connect more than
  two nodes. These are useful for some problems: for instance
  indicating a multi-access medium in a computer network (such as a
  wireless network).

  One can realise hyper-graphs using existing graph representations by
  adding a new type of node (representing the hyper-edge) and creating 
  simple edges from this to all the hyper-edge adjacencies (which can
  then be represented by a node-neighbour or adjacency list for a
  bipartite graph); or by creating ``groups'', whose membership
  represents the hyper-edge.  Hence existing formats can often support
  hyper-edges in principle.  However, true support needs specialised
  data for hyper-edges in the software reading or writing the data, so
  unless a format explicitly states it can support these and presents
  the mechanism, we assume it cannot.

  As a point to note, if hyper-graph support is intended to be
  included in a data format, then the list of graph representations is
  expanded to include the means of describing a hyper-edge:
  \begin{LaTeXdescription}

  \item[direct]: The groups/hyper-edges are directly defined by listing
    the set of nodes included in each;

  \item[indirect]: Node definitions include a group-membership
    attribute that defines which nodes are connected by the defined
    group;

  \item[hmatrix]: A $\{0,1\}$ matrix of size $N \times E$ (where there
    are $N$ nodes and $E$ hyper-edges) maps nodes to hyper-edges. A
    sparse {\em shmatrix} version of this could be stored.

  \end{LaTeXdescription}
  These representations are illustrated in
  \autoref{fig:hyper_graph}. As before none is universally superior,
  though the direct method seems likely to win for most realistic
  graphs. 

\begin{figure*}[htbp]
  \centering
  \rowcolors{2}{white}{white}
  \begin{minipage}[t]{0.3\linewidth}
    \includegraphics[width=\textwidth]{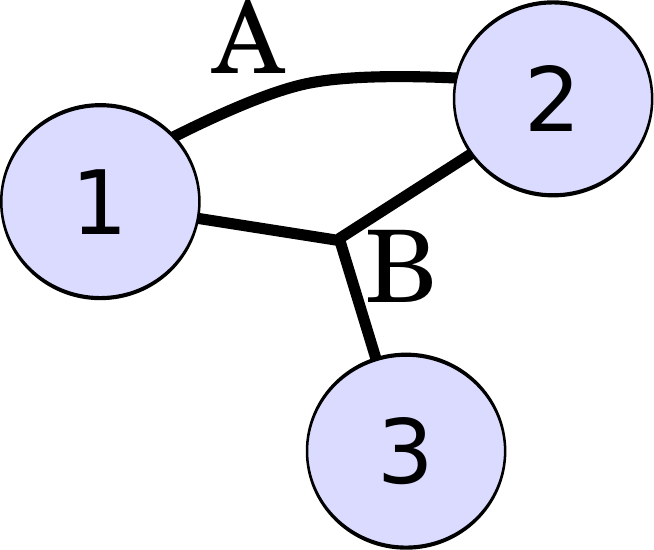}
  \end{minipage}
  \\[3mm]

  \setlength{\currentfiglenth}{0.24\linewidth}
  \begin{minipage}[b]{\currentfiglenth}
    \centering
     \[
      \begin{array}{ll}
        A: & 1,2 \\
        B: & 1,2,3 \\
      \end{array}
      \]
    (a) Direct
  \end{minipage}
  \begin{minipage}[b]{\currentfiglenth}
    \centering
     \[
      \begin{array}{ll}
        1: & A, B \\
        2: & A, B  \\
        3: & B
      \end{array}
      \]

     (b) Indirect.
  \end{minipage}
  \begin{minipage}[b]{\currentfiglenth}
    \centering
     \[ H = \left(
      \begin{array}{ll}
        1 & 1 \\
        1 & 1 \\
        0 & 1 \\
      \end{array}
      \right) 
      \]

      (c) H matrix.
  \end{minipage}

  \caption{A hyper-graph with representations.}
  \label{fig:hyper_graph}
\end{figure*}

\item[hierarchy]: It is common for graphs to have sub-structure, for
  instance nodes that themselves contain graphs.

  Several formats provide mechanisms to record this substructure.
  Unfortunately, there does not seem to be a consistently used
  definition of this type of structure \cite{bildhauer11:_dhhtg}, and
  so we see differences not just in the representation, but also what
  exactly is being represented.  The problem becomes even more
  complicated when hierarchy and hyper-graphs are combined
  \cite{kivela:_multil} (there is at least one proposed solution
  \cite{bildhauer11:_dhhtg} but it does not seem to be widely used
  yet).

  Here, we simply note whether the format provides a version of
  hierarchy.

\item[meta-graph]: A meta-graph~\cite{basu07:_metag_applic} is a
  generalisation of a graph, multi-graph, hyper-graph, and
  hierarchical graph. Once again, a meta-graph could in principle be
  represented using existing data structures (in much the same way
  that any data can in principle be represented in XML), so this
  fields refers to whether the format defines the representation.  As
  far as we know, no format yet supports meta-graphs\footnote{ Note
    the term ``meta-graph'' is somewhat overloaded, \eg there is at
    least one package called {\em metagraph} that has nothing to do
    with the mathematical meta-graph.}, but this is included as a
  feature as an indication of the type of feature that might require a
  new format, or extended version of an existing format.

\item[edge-edge links]: Generally, a graph has links between nodes,
  but we could generalise the concept to allow meta-edges that join
  edges as well (this is different from a meta-graph).

  This idea isn't supported by many formats, and in the case of
  GraphML it is specified using the extensibility of GraphML, but
  again it is a useful example of the types of features that may be
  needed in the future.

\end{LaTeXdescription}
\autoref{tab:graph_types} shows which graph types are supported by
each format.  

\begin{table*}[p]
  \hyphenpenalty=100000
  \centering
  {\footnotesize
    \begin{tabular}{r|llllll}
      \input{excel2latex/tab4.tex}
      \hline
      \input{excel2latex/tab5.tex}  
    \end{tabular} 
  }
  \caption{Graph types (see \autoref{sec:graph_types} for explanation of columns).}
  \label{tab:graph_types}
\end{table*}

\subsection{Attributes}
\label{sec:attributes}

\begin{LaTeXdescription}

\item[edge weights]: A very common requirement is to store a numerical
  value associated with an edge. Generically, we call this a
  weight. Many formats provide the facility to keep one such value.

\item[multiple attributes]: Some formats allow one to keep multiple
  labels (numerical or otherwise) for each node and/or edge. 

  For some formats these are fixed (\eg they allow a name and a
  value), whereas others allow arbitrary lists of attributes. 

\begin{figure}[htbp] 
  \centering
  \vspace{-10mm}
  \includegraphics[width=0.95\columnwidth]{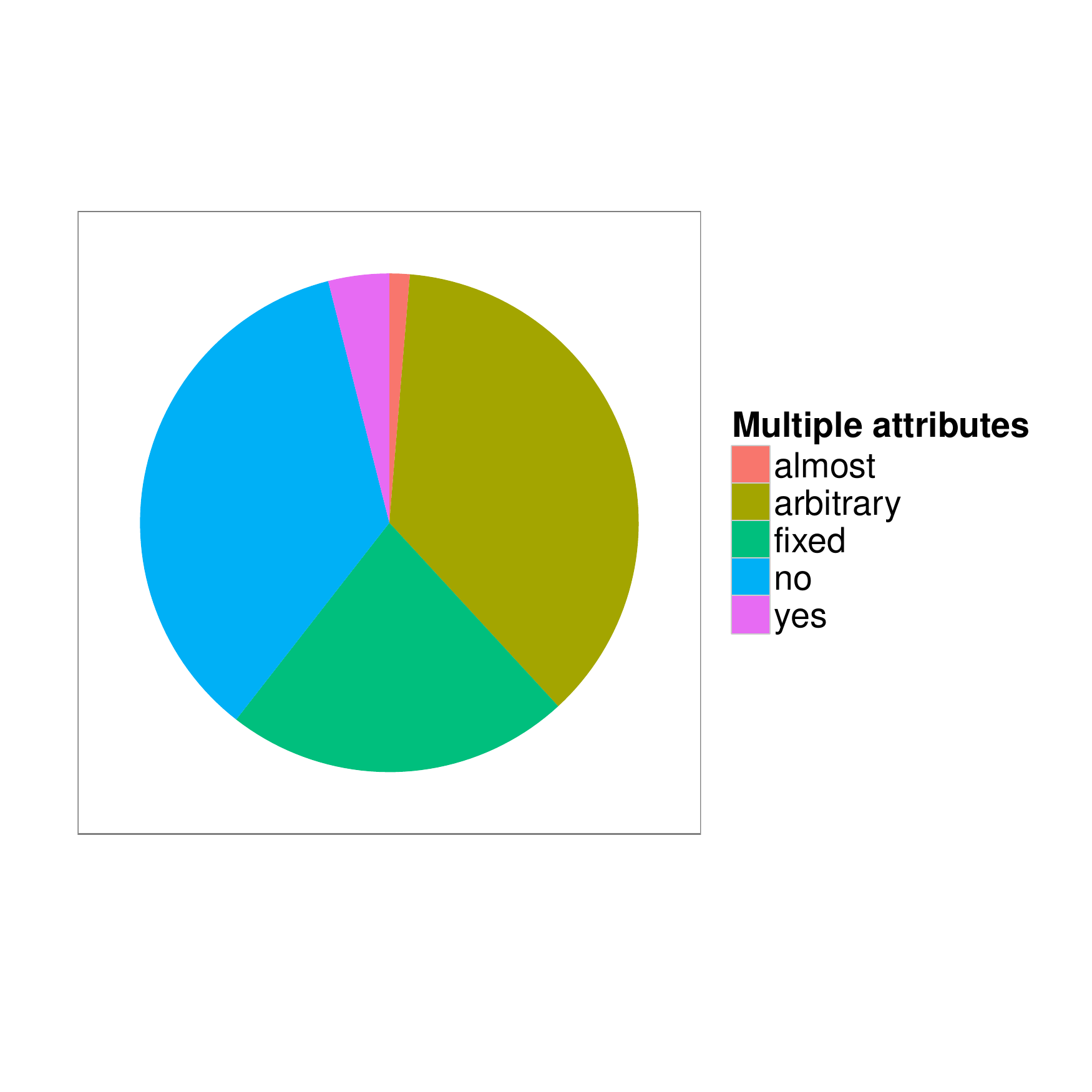}
  \vspace{-16mm}
  \caption{Multiple attribute support.}
  \label{fig:multi_attr_prop}
\end{figure}

\item[default values]: Specifying the value of a weight or attribute
  for every edge or node can be laborious (if it has to be done by
  hand), and wasteful of space. Moreover, it makes it hard to see
  structure in the data. Simply providing a default value for the
  common case can improve the situation. We include here the case of
  simple inheritance of values through a tree of ``class'' structures
  on the objects. For instance, nodes can be given a type which
  conveys a default value to be overridden by a more specific type or
  particular value. Notice here we are not speaking of inheritance
  through the graph itself, but a structure on top of the graph.

\item[multiple inheritance]: A few formats allow values to be derived
  through inheritance of values from multiple classes to which they
  belong. Thus they allow a node to have, for instance, a type
  ``router'' which conveys that it is an Internet router, with
  appropriate characteristics for such a device, from vendor ``Cisco''
  which appropriate characteristics for that vendor.
  
  Once again, inheritance is not through the structure of the graph,
  but through a further structure defined on the graph objects. 

\item[visualisation data]: Files that allow arbitrary attributes can
  always provide data to be used in visualising the graph, but here we
  refer to formats that explicitly provide such data.

  The level of visualisation data varies dramatically: some formats
  only allow position information for nodes, whereas others allow SVG
  definitions to be used in drawing the nodes. Still others provide
  guidance about which layout algorithms to use in displaying the
  graph. 
 
  There is not space here to document all of the variations possible,
  so we simply indicate whether any such data is defined or not. 

\item[ports]: These are a specialised piece of layout information:
  often ports\footnote{Ports are also called hooks in Pajek.} are
  often specified by a compass direction, and indicate where on a node
  the link should join to it. We include ports in addition to the
  previous field because port-based information can also carry
  semantic information about the relationship between links on a
  complex node: \eg the arrangement of links on a real device like an
  Internet router.

\item[temporal data/dynamics]: A topic of interest is
  analysis/visualisation of graphs as they
  change~\cite{ebert87:_versat_data}.  One way to store this
  information is as a series of ``snap-shot'' graphs, but storing it
  all together in the same file has some appeal. A few formats provide
  some variant on this: allowing links or nodes to be given a
  lifetime, or proving ``edits'' to the graph at specific epochs.

\end{LaTeXdescription}
\autoref{tab:attributes} explains the attribute features that are
supported by each format.

\begin{table*}[p]
  \hyphenpenalty=100000
  \centering
  {\footnotesize
    \begin{tabular}{r|p{15mm}p{20mm}p{15mm}p{20mm}p{20mm}lp{20mm}} 
      \input{excel2latex/tab6.tex}
      \hline
      \input{excel2latex/tab7.tex} 
    \end{tabular}
  } 
  \caption{Allowed attributes (see \autoref{sec:attributes} for explanation of columns).}
  \label{tab:attributes}
\end{table*}

\subsection{General}
\label{sec:general}

\begin{LaTeXdescription}
\item[extensible]: Some formats allow extensibility in varying
  forms. We only consider them to have this facility, however, if they
  provide an explicit mechanism. For instance, we do not regard all XML
  derivatives as intrinsically extensible because they could, in
  principle, be extended using standard XML techniques. The format has
  to explain the explicit mechanism whereby it is extended.

  Simply adding extra attributes is not considered extensibility.

\item[schema checking]: A format that provides an explicit mechanism
  to check that a file is in a valid format is useful. We only say it
  has this facility if a tool exists to perform the check (a
  schema-checking program, DTD, or other similar formal tool).

\item[checksums]: It is possible for large data files to become
  corrupted. A common preventative (or at least check for this
  problem) is to use a checksum. This is possible for all files, but
  we say that a given format has this capability if it includes it as
  a internal component (usually checking everything except the
  checksum itself). Only a few formats contain this check.

\item[external data references]: Some formats allow reference to
  external files. This could be for visualisation data, meta-data, or
  other purposes. There are several approaches and views on external
  references, but we record whether it is expected that all relevant
  information will be in the file, or whether there might be something
  external.  Again, we look for an explicit explanation of the
  mechanism, not implicit inheritance from the parent file format.

\item[multiple graphs]: Some formats allow multiple graphs to be held
  in one file. Again, we only count this as a feature if the
  specification explains how explicitly.

\item[incremental specification]: A small number of formats that
  present multiple graphs allow these graphs to be specified
  incrementally. This is subtly different from including temporal
  dynamics, as there is no implication of time, and the different
  graphs could potentially be unrelated (for instance, this might be
  used to describe graph edit distance problems).

  In a sense incremental specification is a simple case of
  constructive graph definition, but it is a very limited case, with
  specific application, so we list it separately.

\end{LaTeXdescription}
\autoref{tab:other} provides information on the other features of the
file formats.

\begin{table*}[p]
  \hyphenpenalty=100000
  \centering
  {\footnotesize
    \begin{tabular}{r|lp{15mm}lp{20mm}p{15mm}p{20mm}}
      \input{excel2latex/tab8.tex}
      \hline
      \input{excel2latex/tab9.tex} 
    \end{tabular} 
  }
  \caption{Other properties (see \autoref{sec:general} for explanation of columns).}
  \label{tab:other}
\end{table*}

\section{Data statistics}

\begin{figure*}[htbp]
	\centering
	\includegraphics[width=0.95\textwidth]{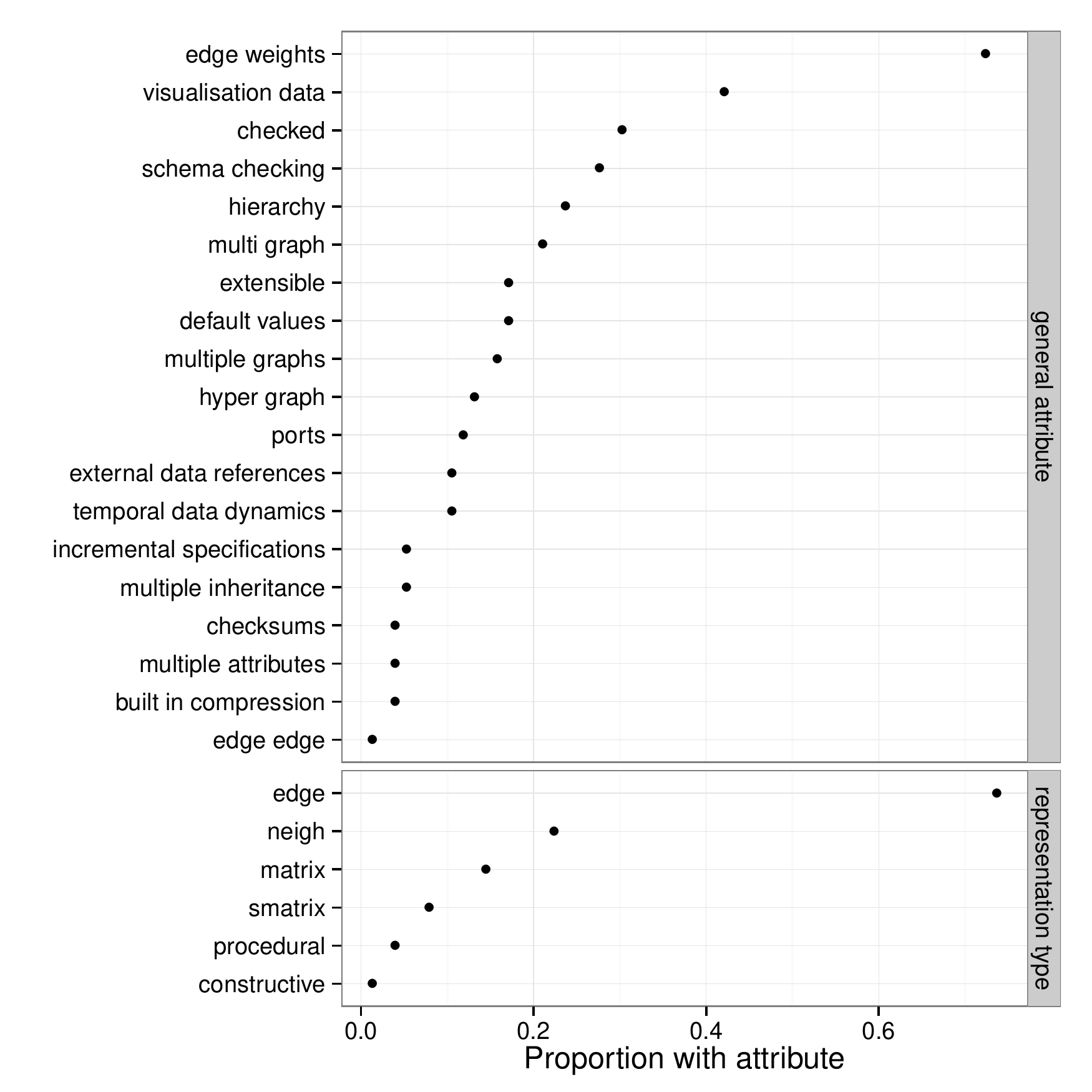}
	\caption{Proportion of each feature in the data formats.}
	\label{fig:yes_plot}
\end{figure*}

In this section we statistically summarise the necessarily large
tables presented earlier. Some of the charts already presented provide
some details, but we explore in more detail by looking at the others
to calculate the proportion of formats supporting each of the features
listed. This is plotted in \autoref{fig:yes_plot}. Note that in regard
to features with multiple answers (\eg representation), we break the
possibilities into categories and list the proportion that support
each category.

Most obviously, there is a large support for edge representations
along with an edge weight. Visualisation data is also widely
supported.  

\begin{figure*}[htbp]
	\centering
	\includegraphics[width=0.9\textwidth]{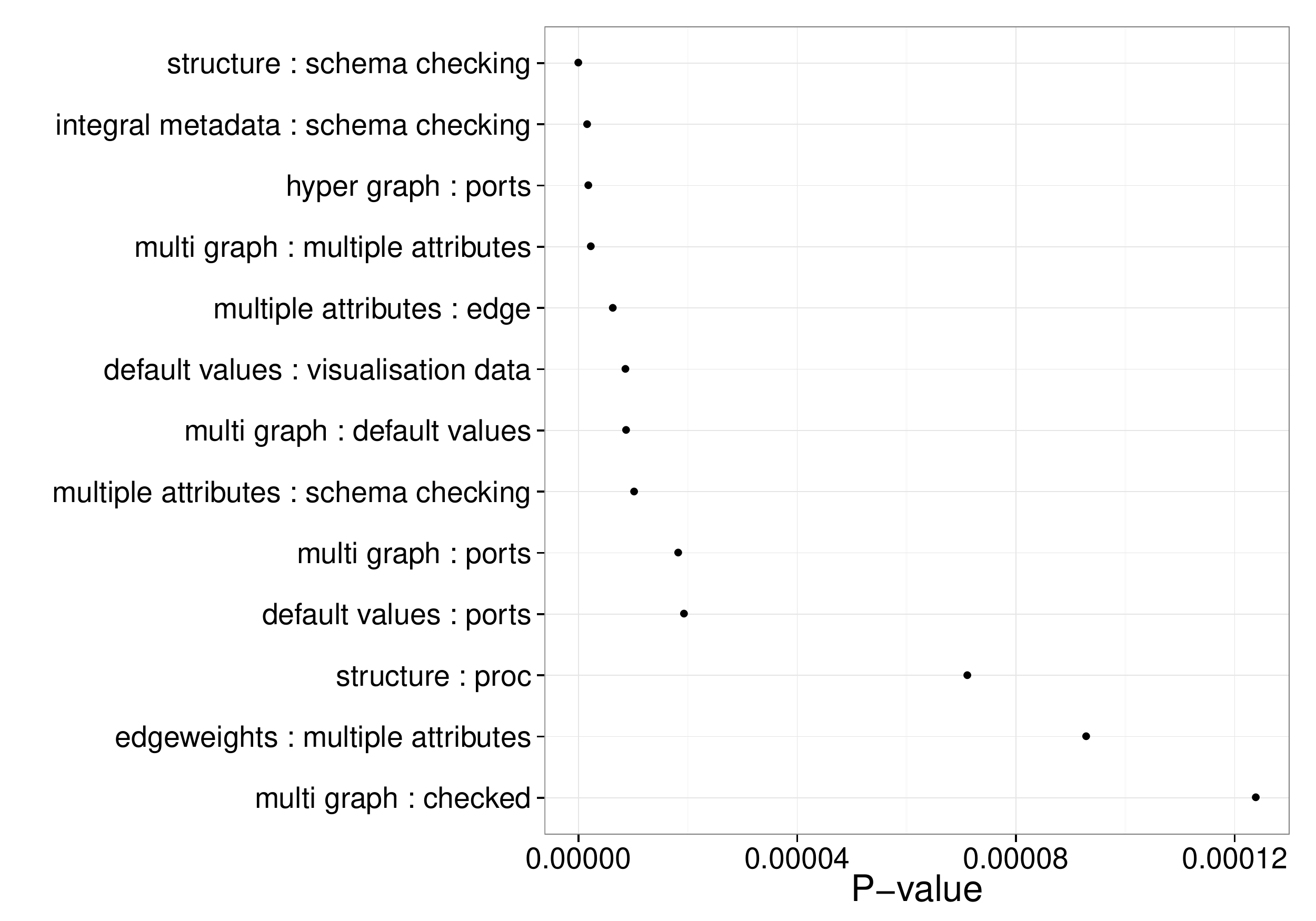}
	\caption{P-values for significant associations between columns.}
	\label{fig:fisher}
\end{figure*}

Next we look at bivariate correlation between columns in the
tables. For each pair of columns, we calculate a contingency table and
then a P-value for the Fisher exact test \cite{Agresti2002}, which is
used because we have lots of small strata. \autoref{fig:fisher} shows
the {\em significantly} correlated pairs, where this is defined as
having a significant P-value after Bonferroni adjustment
\cite{Casella2002}.

Many of the results are obvious. For instance, it is hardly
surprising that there should be a significant correlation between
the file structure and schema checking. 

On the other hand there are many surprising effects:
\begin{itemize}

\item hyper-graphs and ports are often associated; and

\item multi-graph and default values are also associated,

\end{itemize}
These seem to be indications that the type of file author who thinks
carefully about certain aspects of the file (\eg the types of graphs
that will be represented) also thinks about other aspects that require
care. Thus dividing the formats in ``careful'' and ``quick and
dirty''. More work is required to establish if this connection is
genuine or merely accidental. 

\vspace{15mm}

\section{Decisions}

The list above is not intended to be pejorative. However, it is
potential users need to make decisions about which format to
use. There are several issues that need be considered in such a
decision, and although the first is the feature list required, there
are others:
\begin{LaTeXdescription}

\item[data size]: The size of the graph data to be recorded and used
  is an important factor in file format decisions. This is sometimes
  glossed over when XML-style formats are considered: these are very
  redundant formats, and hence much larger than needed, but they
  compress well. Hence, the compressed version may be no longer than a
  tighter initial specification. However, the issue of read/write time
  (and indeed compression/decompression time) still depends greatly on
  the format's wordiness. Large graphs need tighter formats: either
  binary formats, or at least those that avoid unnecessary bloat. 

  On the far end of the spectrum is the possibility of graph-specific
  compression being part of the storage process (much as many image
  formats provide image compression as an integral features). Only one
  format we found provides true graph-based compression: BVGraph. 

\item[edge density]: Edge density affects the choice of best
  representation of a graph. Very sparse graphs are best represented
  by edge lists, moderately sparse graphs are (perhaps) slightly
  better stored as neighbour lists, and dense graphs may be better
  stored as a full adjacency matrix. 

\item[access method]: Most graph formats are designed to be read
  serially directly into memory in their entirety. Only BVGraph seems
  to provided support for random (or indexed subgraph) access to part
  of a graph. 

  Another example of alternative access methods is that many graph
  algorithms can be reduced to a generalised matrix-vector product,
  and can be performed by repeatedly streaming the edges from disk
  without loading the graph into memory, which is necessary if the
  data is truly large \cite{gleich14}. 

  Further, formats could potentially enable reading the graph in
  parallel to exploit clustered computing\cite{gleich14}. 

  In other cases, a single graph might be part of a larger database.

  In general these issues seem to have been left in the field of graph
  databases \cite{Angles:2008}, and not considered for exchange of
  data.

\item[human readability]: Portability requires the file to be machine
  readable, but a file that is more easily understood by humans is
  potentially better because it is easier to enter and check. Many of
  graph examples datasets were entered at least in part by hand: often
  through a spreadsheet or text editor, and are maintained in the same
  way.  In the case of the Internet topology Zoo~\cite{Zoo} the data
  were entered ``semi-manually'' through yED (a graph editing
  program).
 
  Human readability requires a text file in a logical format, but it
  also needs to avoid: (i) bloat, which distracts the reader with
  unnecessary text, and (ii) the file to be organised neatly. XML
  formats often fail on these: the first because of the volume of
  tags, and the second because they allow organisations which are
  unreadable, \eg with all the text on one line.

  Ultimately, human readability is a highly subjective criteria. Some
  people may find XML easy to read, and others get distracted by the
  tags. As such, we won't comment on it further here.
 
\item[maintenance]: The document
  \cite{proteom_stand_initiat_molec_inter} deals with the use cases
  for graphs, which we can broadly classify (in simpler nomenclature)
  as
  \begin{LaTeXdescription}
  \item[creator]: originally creates the data set, 
  \item[investigator]: uses the data for some purpose, and
  \item[curator]: refines and corrects the data.
  \end{LaTeXdescription}
  Most current graph-exchange formats are oriented at creation and
  investigation, but not curation.

  Data can easily contain errors, and correcting these {\em ex post
    facto} should be supported, but most formats do not deal with
  issues such as
  \begin{LaTeXdescription}
  \item[version control]: to allow, for instance, users to know
    exactly which dataset was used in a particular publication; and
  \item[diff]: the ability to find semantic differences between two
    files to learn what changed between the two (as opposed to simply
    seeing syntactical differences).
  \end{LaTeXdescription}
  Taking differences of arbitrary graph data is hard (it involves
  solving the near-isomorphism problem), but much graph data is
  labelled and in this case differences can be found easily.

\item[documentation]: Through compiling the information used in this
  paper it has become obvious that a key limitation of many formats is
  incomplete documentation. Hidden assumptions, specification by
  (limited) examples, and/or documentation by source code are all
  common. Ideally, any truly portable format should have a complete,
  highly-specific schema; human readable documentation (with
  examples); and source code. All of these together provide the
  ideal documentation.

\item[support]: Finally, the support for the format in a variety of
  tools is a crucial requirement for exchange of data. Likewise,
  support for formats in a variety of public databases makes it more
  useful. We shall consider this issue in more detail below.   

\end{LaTeXdescription}

\subsection{Software Support}

The most difficult issue surrounding software support is that a piece
of software may notionally support a file format, and yet still be
incompatible with other software notionally supporting the same
format. 

For instance, software might
\begin{itemize}
\item fail to accept integers outside a particular range;
\item have varying case sensitivity; 
\item be unable to read the right character set;
\item be unable to read strings beyond a particular length (very few
  formats specify buffer or string lengths); or
\item fail to cope with files larger than some size. 
\end{itemize}
Size is interesting, because almost no documentation exists for size
limits for any data formats. However, it should be reasonably obvious
that if 32 bit integers are used, then the largest number of (integer)
identifiers is around 4 billion. In the past this was large enough
that the need to specify it may have seemed small. With today's
graphs, this could be an important limitation. 

Even more pernicious is partial support for a format. Even when
documented this makes our job hard, but partial support is not often
documented. Instances include:
\begin{itemize}
\item hyper-graphs supported in the format, but not in software;  or
\item some small  number of formats make mention of allowing complex
  numbers; or
\item partial support for hierarchy (\ie the file can be read, but
  the subgraph structure is not retained).
\end{itemize}

Even more complex is the fact that some features may be supported on
read or write, but not both. 





The list of potential software is long, even more so than the list of
formats, so we won't try to survey them here as well. Instead we refer
readers to \cite{bodlaj13:_networ}, which contains a cross-section of
both formats and their software support. 

A common conclusion amongst those who look at this type of data is
that GraphML and Pajek are the most commonly supported in modern
systems, but they are by no means universal or even supported by the
majority of tools.

Another related issue is how hard it would be to provide support
for a format in a new tool. This is a complex issue, but there are
several factors that influence it. Documentation, as mentioned above,
is a critical issue, as is the ability to use existing tool-sets such
as those for XML or JSON. However, one issue hasn't been discussed,
which is the provision of an adequate test dataset.
  
\subsubsection{Test cases}

It's a tautology that implementation of a new graph format isn't
terribly hard, except for the hard bits. The point is, though, that
many formats don't tackle these. 

Many areas of difficulty are listed above. One we have not discussed
in detail is the existence of test cases.  Ideally, in addition to a
complete specification, there should be a set of accompanying files
providing encoded data to demonstrate each feature over a reasonable
range of values \cite{bender-demoll14}.  These files could then be
used by other developers to check their parser implementations. 

The concept of test cases is from software engineering 101.  However,
we are not aware of a single format that provides a truly complete
set. Some provide a set of small examples, but these don't
express all of the features of the data. For instance, encoding, size
limits, advanced features and so on are rarely considered in these
examples. Other formats are used for exchange of datasets, and these
form a {\em de facto} 

More often, only a small set of examples is provided, and these don't
express all of the features of the data. For instance, encoding, size
limits, advanced features and so on are rarely considered in these
examples. Other exchange sets are used to provide large datasets, but
these two are unsuitable for test purposes because they are large and
complex, and don't exercise features in isolation. What is needed, is
a set of test cases that exercise the features in a controlled and
testable manner.

\subsection{Public DB support}

The other type of support we might wish to see is general support
amongst those who provide data publicly. There are many public
databases that provide example networks for benchmarking or
research. We provide a list in \autoref{tab:pdb} of some of the better
known of these with their format choices. Additional data sources are
listed in \cite{alhajj14:_sourc_networ_data}, and a detailed taxonomy
and examples of computer-network data appears in
\cite{battista13:_handb_graph_drawin_visual}.

\begin{table*}[ptb]
  \centering
  {\small
    \begin{tabular}{rl|p{50mm}p{50mm}}
      \input{excel2latex/tab10.tex}
      \hline
      \input{excel2latex/tab11.tex} 
    \end{tabular} 
  }  
  \caption{Public Databases. NB: there is some overlap in the data
    kept in these repositories.}
  \label{tab:pdb}
\end{table*}

There is no clear winner here: slightly preferred is a variant of the
Trivial Graph Format due to its least-common-denominator status (but
note that this isn't really one format, so much as a collection of
equivalent formats). Overall, however, the formats seem to be written
for the data rather than the other way around. That, in itself, is an
illustration of the problem.

\subsection{Future considerations}

There are many considerations or features that we could consider. The
set chosen above were chosen for the illustrative value, given current
graph exchange concerns. In the future, there are other features that
could become interesting, and we list and discuss some of these in the
following. In general, we have not tried to classify the file formats
by these features simply because it seems that few formats support
these, but information is sparse and it is difficult to be certain in
many cases. Many of the issues cross over into issues that {\em have}
been considered in the domain of graph databases \cite{Angles:2008},
and so techniques to tackle the problems exist, but have not been
applied to the world of exchanging data. We will discuss at least a
few of these issues below.

 \begin{LaTeXdescription}

 \item[self-describing]: this refers to whether a file provides its
   own definition of its format. XML arguably has this property, but
   still relies on correct semantic interpretation of arbitrary
   labels, for instance link ``weight'' could mean several different
   things, and have any number of different units. 

 \item[data distribution]: most graph formats are monolithic in that
   the entire graph is held in one file. Even those that allow
   multiple files use this to structure the type of information each
   contains, not to spread the information evenly.

   As graph data becomes larger, and the need to query subsections of
   the graph grows, we need to be able to create modularity in the
   graph representation. Formats that provide the ability to distribute
   the graph information over multiple (indexed) files provides a
   capability that could be very useful \cite{bildhauer11:_dhhtg}.

   This type of consideration, however, seems to have been limited
   primarily to graph databases \cite{Angles:2008}, not exchange
   formats.

 \item[node list]: does the format have a separate node list, or is
   this list implicit in the edges?

 \item[multi-layer]: generalisations of graphs can have a layer
   structure~\cite{kivela:_multil} (resembling in some cases
   hierarchy, and in some cases temporal evolution, but more flexible
   than either by itself). Multi-layer graphs can naturally be
   described by adjacency tensors, however, complete multi-layer
   support doesn't yet appear in any format. 

 \item[linear indexing]: Another consideration in classifying network
   graph formats in the future is whether they use {\em linear
     indices} \cite{gleich15}, by which we mean that if the network
   has $n$ nodes, then they are labelled $1,2,\ldots, n$ (equivalently
   we could start at 0).

   Linear indexes make a dataset easier to deal with at two
   levels. Firstly, it is more efficient to store integers than
   arbitrary strings: so both node and edge lists can be read/written
   more quickly, but also when the data is read into a program if the
   node names are arbitrary then the node data needs an extra layer of
   indirection such as provided by an associative array, and for large
   datasets this can reduce performance compared to storing the data
   in a simply indexed array. 

   The issue is primarily important for very large datasets, but these
   are becoming more common.

   Note that it doesn't mean that nodes can't be named: they can have
   all the usual meta-data one might associate with the node, but it
   means that the primary reference to the node is arithmetically
   simple to work with.
   
   In general, matrix representations have an implicit linear
   indexing, but other formats are less clear about the
   issue. However, some illustrative examples include SNAP, which uses
   integer but not linear indices and Matlab and BVGraph, which both
   use linear indices.

   One can also imagine creating simple indexes into the edges, but
   this goes a step beyond any exchange formats goals so far.

 \item[serialisation]: Many graph data formats are designed to be read
   into memory in their entirety. They do not support the ability to
   read through the data serially, and perform analysis on the fly. 

 \item[random access and/or queries]: As noted, most graph data
   formats are designed to be read into memory in their entirety.

   But an even bigger limitation, even of those that can be read
   serially is that they do not support the ability to find
   information about an arbitrary link or node (or subset of such)
   without reading the whole data set (at least through to the
   relevant point). 

   Again, this is only a problem for very large datasets, but clearly
   is a huge issue for such sets. Not least because it is easy to
   imagine datasets to large to be read into realistic RAMs, but also
   because this is hopelessly inefficient for certain types of
   analysis.

   Again, graph databases deal with this issue, but exchange formats
   have not, so far.

\item[parallel read/write]: The monolithic nature of most graph
  exchange formats make them unsuitable for parallel writing. It is
  hard to separate parts of a graph and write them independently. 

  The fact that it is assumed that most files will be read in their
  entirety also limits the ability to parallelise read operations. 

  Again graph databases attack this problem, whereas exchange formats
  have not. 

 \end{LaTeXdescription}

\subsection{Discussion}

The point of all this: what should be done here, how should one
proceed. There are three major considerations:
\begin{itemize}
\item what representation of a graph (or generalised graph) will be
  used: edge or neighbour list, adjacency matrix, paths, or some
  constructive or procedural approach;  
\item what additional information is to be added, and how flexible
  this information should be; and
\item what encapsulation of the data is to be used (XML and more
  recently JSON seem to be favourites). 
\end{itemize}
Then there are a substantial set of other features and aspects of the
dataset that should be considered in the choice of formats. 


 




 

\section{Conclusion}

The science of graphs and networks needs portable, well-documented,
precisely-defined, exchange formats. There are many existing formats,
and this paper seeks to unravel this mess, most notably with the aim
of reducing the number of new formats developed.

One size probably does not fit all though.  There is a clear need for
at least three major types of file format:
\begin{itemize}

\item a general, flexible, extensible approach such as GraphML;

\item a quick and dirty approach that satisfies the least common
  denominator for the exchange of information to/from the simplest
  software; and

\item a very efficient (compressed) format for very large graphs. 

\end{itemize}

Its not clear that any format at present has a complete enough list of
features to take the roll of the first format. No doubt this will
continue to evolve as well, as new features are required. Moreover,
the requirement human readability of the data is evolving as more
datasets are generated through automated means rather than entered by
hand. 

The second is easy, but there are very many contenders, and settling
on one will be hard.
 
The final one should be seen as an interesting research topic given
there are multiple compression techniques available. However, the only
true example of a compressive format is BVGraph does not allow
attributes, and so some thought might be devoted to that topic. 

Finally, although having arbitrarily extendable attributes for the
graph and its components seems an attractive feature, it is easy to
see why specialised applications would prefer a pre-defined list. Most
obviously to make support for those attributes easier (both in terms
of parsing\footnote{The ability to specify arbitrary attributes
  usually comes at the cost of a more complex mechanism being required
  to read and write these. The cost is usually in terms of supporting
  code complexity, and read/write times.}, and in terms of
exchange\footnote{Exchange requires common definitions of the
  meaning of the attributes, not just syntax. If the attributes are
  arbitrary then some information might be mistranslated by use of
  different attributes to hold similar information, or the same
  attribute to hold different information. For example, for graphic
  attributes, including the size of a vertex to be drawn is not very
  useful without well-defined units.}). However, there is also the
subtle issue of what attributes {\em could} be included vs those that
{\em should} be included. Explicit definition of the {\em required}
attributes can create a better overall set of data by forcing the
lowest-common-denominator to be higher. 
 
In the end, maybe what is needed is actually a container format:
allowing specification of parts of a graph in alternative formats. Or
allowing specification of meta-data and labels in an XML-like format,
but the edge data in a more compact form. 

Alternatively, good conversion programs could simplify the issue, but
at present most software tools are not designed with this in mind (for
instance, such a tool needs to be lightweight, but warn about
different available features, and support a large range of
possibilities).
  
\section*{Acknowledgements}

This work was supported by ARC grant DP110103505, and by the ARC
Centre of Excellence for Mathematical \& Statistical Frontiers.

Many people have contributed to improve the quality of information
presented here; specific thanks go to 
Andreas Winter, 
Andy Sch\"{u}rr,
Brendan McKay, 
Danny Bickson, 
Ivan Herman, 
Kevin Kawkins, 
Mason Porter, 
Michael Himsolt, 
Peter Mucha, 
Rok Sosic,  
Sebastian Mueller,
Skye Bender-deMoll, 
Syd Bauman, 
S\'{e}bastien Heymann, 
Tels, 
Ulrik Brandes, 
Vladimir Batagelj, 
Bruce Hendrickson, 
David Gleich,  
Young Hyun,
Antoine Dutot,
Rose Oughtred (and the BioGRID Administration Team),  and
David Krackhardt. 



{\small  
\bibliographystyle{IEEEtran}
\bibliography{topology,ip_traffic,books}
}


\end{document}

%% file: excel2latex/tab0.tex
 & {\bf {Graph Format}} & & & {\bf {Full Name}} & \multicolumn{2}{p{18mm}}{{\bf {Reference time frame} {}}}\\ 

%% file: excel2latex/tab1.tex
{1} &  \href{http://docs.graphlab.org/graph_formats.html}{bintsv4} & \cite{graphlab} & {} & {bintsv4 (GraphLab)} & {2009}\; & {present} \\ 
{2} &  \href{http://wiki.thebiogrid.org/doku.php/biogrid_tab_version_2.0}{BioGRID TAB} & \cite{chatr-aryamontri12:_biogr_inter_datab,biogr } & {\checkmark } & {BioGRID TAB 2.0 Format} & {2003}\; & {present} \\ 
{3} &  \href{http://www.dia.uniroma3.it/~gdt/gdt4/blag.php}{BLAG, GDToolkit} & \cite{gdtoolkit} & {} & {Batch layout generator (GDToolkit)} & {1998}\; & {2008} \\ 
{4} &  \href{http://webgraph.di.unimi.it/docs/it/unimi/dsi/webgraph/BVGraph.html}{BVGraph} & \cite{boldi04:_bvgraph} & {\checkmark } & {Boldi-Vigna graph compression} & {2004}\; & {2011} \\ 
{5} &  \href{http://www.sandia.gov/~bahendr/papers/guide.ps}{Chaco} & \cite{hendrickson95:_chaco} & {\checkmark } & {Chaco graph format} & {1994}\; & {1995} \\ 
{6} &  \href{http://glaros.dtc.umn.edu/gkhome/cluto/cluto/overview}{Cluto} & \cite{karypis03:_cluto_clust_toolk} & {} & {Cluto/Metis/Graclus format} & {1999}\; & {2008} \\ 
{7} &  \href{http://graphstream-project.org/doc/Advanced-Concepts/The-DGS-File-Format_1.1/}{DGS} & \cite{dgs_file_format_specif} & {\checkmark } & {Dynamic GraphStream Format} & {2010}\; & {2013} \\ 
{8} &  \href{http://schemas.microsoft.com/vs/2009/dgml/}{DGML} & \cite{levinson10:_dgml} & {} & {Directed Graph Markup Language} & {2009}\; & {2013} \\ 
{9} &  \href{http://www.dis.uniroma1.it/challenge9/format.shtml}{DIMACS} & \cite{dimacs9} & {} & {DIMACS graph format} & {2006}\; & {2006} \\ 
{10} &  \href{http://www.graphviz.org/content/dot-language}{Dot} & \cite{mutzel02:_graph} & {} & {GraphVis Dot Language} & {2000}\; & {present} \\ 
{11} &  \href{http://martin-loetzsch.de/DOTML/}{DotML} & \cite{loetzsch:_dotml} & {} & {Dot Markup Language} & {2002}\; & {2010} \\ 
{12} &  \href{http://www.casos.cs.cmu.edu/projects/dynetml/}{DyNetML} & \cite{tsvetovat04:_dynet} & {} & {DyNetML XML} & {2001}\; & {2009} \\ 
{13} &  \href{http://www.nas.nasa.gov/Software/GAMFF/gamffpaper.ps}{GAMFF} & \cite{zien95:_gamff} & {} & {A Graph and Matrix Format} & {1995}\; & {1995} \\ 
{14} &  \href{http://guess.wikispot.org/The_GUESS_.gdf_format}{GDF} & \cite{guess} & {} & {Guess Data Format} & {2007}\; & {2010} \\ 
{15} &  \href{http://netwiki.amath.unc.edu/DataFormats/GDL, vcgdoc.ps.gz}{GDL} & \cite{sander95:_gdl} & {} & {Graph Description Language} & {1993}\; & {1995} \\ 
{16} &  \href{http://homepages.rootsweb.ancestry.com/~pmcbride/gedcom/55gctoc.htm}{GEDCOM} & \cite{96:_gedcom} & {} & {Geneaological data} & {1987}\; & {1996} \\ 
{17} &  \href{http://gexf.net/format/}{GEXF} & \cite{heymann:_gexf} & {\checkmark } & {Graph Exchange XML Format} & {2007}\; & {2012} \\ 
{18} &  \href{http://www.fim.uni-passau.de/index.php?id=17297&amp;L=1}{GML} & \cite{himsolt95:_gml} & {\checkmark } & {Graph Modelling Language} & {1995}\; & {1999} \\ 
{19} &  \href{http://cs.anu.edu.au/~bdm/data/formats.txt}{Graph6} & \cite{brinkman:_guide} & {\checkmark } & {Graph6} & {1996}\; & {2011} \\ 
{20} &  \href{http://bloodgate.com/perl/graph/manual/}{Graph::Easy} & \cite{tels07:_grapheasy} & {\checkmark } & {Perl Graph::Easy format} & {2004}\; & {present} \\ 
{21} &  \href{http://www3.cs.stonybrook.edu/~algorith/implement/graphed/distrib/README}{GraphEd} & \cite{graphed,Himsolt_graphed:a} & {} & {GraphEd simple format} & {1994}\; & {1994} \\ 
{22} &  \href{http://www.graphjson.org/}{GraphJSON} & \cite{graphjson} & {} & {Graph JSON} & {2013}\; & {2014} \\ 
{23} &  \href{http://graphml.graphdrawing.org/}{GraphML} & \cite{brandes13:_graphml} & {\checkmark } & {Graph Markup Language} & {2000}\; & {present} \\ 
{24} &  \href{https://github.com/tinkerpop/blueprints/wiki/GraphSON-Reader-and-Writer-Library}{GraphSON} & \cite{graphjson} & {} & {TinkerPop's JSON-based Graph format} & {2011}\; & {2013} \\ 
{25} &  \href{http://link.springer.com/chapter/10.10070.0000003-540-44541-2_6}{GraphXML} & \cite{herman02:_graph_drawin} & {\checkmark } & {XML-Based Graph Description Language} & {1998}\; & {1998} \\ 
{26} &  \href{http://www.se.uni-oldenburg.de/documents/ebert+1999.pdf}{GraX} & \cite{ebert99:_grax} & {} & {GraX} & {1999}\; & {1999} \\ 
{27} &  \href{http://www.user.tu-berlin.de/o.runge/tfs/projekte/gxl-gtxl/kent.html}{GRXL} & \cite{rodgers:_grxl} & {} & {XML Specification for Grrr Program} & {2000}\; & {2000} \\ 
{28} &  \href{http://www.cc.gatech.edu/fac/Ellen.Zegura/graphs.html}{GT-ITM} & \cite{Z} & {} & {Georgia Tech Internetwork Topology Models} & {1996}\; & {1998} \\ 
{29} &  \href{http://www.gupro.de/GXL/}{GXL} & \cite{holt06:_gxl} & {\checkmark } & {Graph eXchange Language} & {1999}\; & {2006} \\ 
{30} &  \href{http://people.sc.fsu.edu/~jburkardt/data/hb/hb.html}{Harwell-Boeing} & \cite{duff92:_harwell_boeing} & {} & {Harwell-Boeing sparse (TGFaceny) matri} & {1992}\; & {2010} \\ 
{31} &  \href{http://topology.eecs.umich.edu/inet/}{Inet} & \cite{jin00:_inet} & {} & {Inet Topology Generator file} & {2000}\; & {2002} \\ 
{32} &  \href{http://www.caida.org/data/internet-topology-data-kit/}{ITDK} & \cite{14:itdk} & {\checkmark } & {CAIDA Internet Topology Data Kit} & {2002}\; & {present} \\ 
{33} &  \href{https://github.com/jsongraph/json-graph-specification}{JSON Graph} & \cite{json_graph} & {} & {json-graph-specification} & {2014}\; & {present} \\ 
{34} &  \href{http://www.algorithmic-solutions.info/leda_guide/graphs/leda_native_graph_fileformat.html}{LEDA} & \cite{leda} & {} & {LEDA format} & {2001}\; & {2008} \\ 
{35} &  \href{http://lemon.cs.elte.hu/pub/doc/1.2.3/a00002.html}{LGF} & \cite{lemon_graph_format_lgf,lemon_tutor_lemon_graph_format} & {} & {LEMON Graph Format} & {2008}\; & {present} \\ 
{36} &  \href{http://lgl.sourceforge.net/}{LGL} & \cite{adai04:_lgl,marcotte:_large_graph_layout} & {} & {Large Graph Layout} & {2003}\; & {2005} \\ 
{37} &  \href{http://www.caida.org/tools/visualization/libsea/}{LibSea} & \cite{libsea} & {\checkmark } & {CAIDA LibSea format} & {2000}\; & {2005} \\ 
{38} &  \href{http://www.andrew.cmu.edu/user/krack/documents/krackplot_manual.doc}{KrackPlot} & \cite{krackplot} & {\checkmark } & {KrackPlot data format} & {1993}\; & {present} \\ 
{39} &  \href{http://au.mathworks.com/help/matlab/ref/save.html}{Matlab} & \cite{matlab} & {} & {Matlab saved workspace} & {1996}\; & {present} \\ 
{40} &  \href{http://math.nist.gov/MatrixMarket/formats.html}{Matrix} & \cite{boisvert96:_matrix_market} & {} & {Matrix Market sparse matrix} & {1996}\; & {2013} \\ 
{41} &  \href{http://mivia.unisa.it/datasets/graph-database/arg-database/}{Mivia} & \cite{arg_datab_docum,DeSanto:2003:LDG:763451.763457} & {} & {Mivia ARG database format} & {2001}\; & {2003} \\ 
{42} &  \href{http://www.sfu.ca/personal/archives/richards/Pages/dataform.htm}{MultiNet} & \cite{seary05:_multin,multinet} & {} & {MultiNet} & {1999}\; & {2007} \\ 
{43} &  \href{http://www.analytictech.com/Netdraw/NetdrawGuide.doc}{Netdraw VNA} & \cite{borgatti02:_netdraw,netdraw} & {} & {Netdraw VNA} & {2005}\; & {2008} \\ 
{44} &  \href{http://vlado.fmf.uni-lj.si/pub/networks/netml/default.htm}{NetML} & \cite{batagelj95:_towar_netml} & {} & {Network Markup Language} & {1995}\; & {1995} \\ 
{45} &  \href{http://lgl.sourceforge.net/}{Ncol} & \cite{adai04:_lgl,marcotte:_large_graph_layout} & {} & {Large Graph Layout} & {2003}\; & {2005} \\ 
{46} &  \href{http://wiki.cytoscape.org/Cytoscape_User_Manual/Network_Formats}{NNF} & \cite{cline07:_integ_cytos,cytoscape_user_manual} & {} & {Nested Network Format} & {2003}\; & {present} \\ 
{47} &  \href{http://www.andrew.cmu.edu/user/krack/documents/krackplot_manual.doc}{Nod} & \cite{krackplot} & {} & {KrackPlot Node format} & {1993}\; & {present} \\ 
{48} &  \href{http://svitsrv25.epfl.ch/R-doc/library/sna/html/write.nos.html}{NOS} & \cite{n_o_s_format} & {} & {Neo Org Stat format} & {2000}\; & {2013} \\ 
{49} &  \href{http://www.isi.edu/nsnam/ns/doc/index.html}{ns-tcl} & \cite{ns2,fall11:_vint_projec} & {} & {ns-2 Tcl network definition} & {1989}\; & {2011} \\ 
{50} &  \href{http://ogdl.org/spec/}{OGDL} & \cite{veen14:_ogdl} & {\checkmark } & {Ordered Graph Data Language} & {2002}\; & {present} \\ 
{51} &  \href{http://www.ogdf.net/lib/exe/fetch.php/documentation.pdf}{OGML} & \cite{chimani13:_ogml,ogml} & {} & {Open Graph Markup Language} & {2012}\; & {present} \\ 
{52} &  \href{http://biodata.mshri.on.ca/osprey/OspreyHelp/FileIO.html#sec312}{Osprey} & \cite{osprey} & {} & {Osprey file format} & {2001}\; & {2008} \\ 
{53} &  \href{http://www.caida.org/tools/visualization/otter/paper/}{Otter} & \cite{huffaker99:_otter} & {} & {Otter's native format} & {1999}\; & {1999} \\ 
{54} &  \href{http://pajek.imfm.si/doku.php?id=pajek}{Pajek (.net)} & \cite{pajek_progr_large_networ_analy,v.98:_pajek,pajek} & {\checkmark } & {Pajek Tool's .net format} & {1996}\; & {present} \\ 
{55} &  \href{http://pajek.imfm.si/doku.php?id=pajek}{Pajek (.paj)} & \cite{pajek_progr_large_networ_analy,v.98:_pajek,pajek} & {\checkmark } & {Pajek Tool project (.clu, .vec, .per, ...)} & {1996}\; & {present} \\ 
{56} &  \href{http://cs.anu.edu.au/~bdm/plantri/plantri-guide.txt}{Planar} & \cite{brinkman:_guide} & {\checkmark } & {Plantri Planar Code andedgeCode} & {1996}\; & {2011} \\ 
{57} &  \href{http://psidev.sourceforge.net/molecular_interactions//xml/doc/user/}{PSI MI} & \cite{psimi} & {} & {Protenomics Standards Initiative Molecular Interaction } & {2002}\; & {present} \\ 
{58} &  \href{http://www.program-transformation.org/Transform/RigiRSFSpecification}{RSF} & \cite{Kienle:rigi} & {} & {Rigi Standard Format} & {1999}\; & {2010} \\ 
{59} &  \href{http://www.cs.washington.edu/research/projects/networking/www/rocketfuel/}{Rocketfuel} & \cite{rocketfuel_0} & {} & {Rocketfuel ISP Maps} & {2002}\; & {2003} \\ 
{60} &  \href{https://www.cise.ufl.edu/research/sparse/matrices/DOC/rb.pdf}{Rutherford-Boeing} & \cite{duff97:_ruther_boeing} & {} & {Rutherford-Boeing sparse (TGFaceny) matri} & {1997}\; & {1997} \\ 
{61} &  \href{http://www-cs-faculty.stanford.edu/~uno/sgb.html}{SGB} & \cite{Knuth:1993:SGP:164984,knuth:_stanf_graph} & {} & {Stanford GraphBase} & {1992}\; & {2009} \\ 
{62} &  \href{http://xml.coverpages.org/sgfWWW7.html}{SGF} & \cite{Liechti199811_SGF,sgf} & {} & {Structured Graph Format} & {1998}\; & {1999} \\ 
{63} &  \href{http://martin-loetzsch.de/S-DOT/}{S-Dot} & \cite{loetzsch:_sdot} & {} & {S-Dot (lisp interface to Graphviz)} & {2006}\; & {2010} \\ 
{64} &  \href{http://wiki.cytoscape.org/Cytoscape_User_Manual/Network_Formats}{SIF} & \cite{cline07:_integ_cytos,cytoscape_user_manual} & {} & {Simple Interaction Format} & {2003}\; & {present} \\ 
{65} &  \href{http://snap.stanford.edu/}{SNAP} & \cite{snap} & {\checkmark } & {Stanford Network Analysis Platform} & {2005}\; & {present} \\ 
{66} &  \href{http://sourceforge.net/p/sonia/wiki/Son_format/}{SoNIA} & \cite{bender-demoll06:_sonia,sonia_social_networ_image_animat} & {\checkmark } & {So NIA Son format} & {2002}\; & {present} \\ 
{67} &  \href{http://cs.anu.edu.au/~bdm/data/formats.txt, http://computationalcombinatorics.wordpress.com/2012/08/05/tips-and-tricks-using-gtools/}{Sparse6} & \cite{brinkman:_guide} & {\checkmark } & {Sparse6} & {1996}\; & {2011} \\ 
{68} &  \href{http://www.gmw.rug.nl/~stocnet/content/downloads/Stocnet_Manual_17.pdf}{StOCNET} & \cite{boer06:_stocn} & {} & {StOCNET native format} & {2002}\; & {2007} \\ 
{69} &  \href{http://www.tei-c.org/Vault/P4/doc/html/GD.html}{TEI} & \cite{text_encod_initiat} & {} & {Text Encoding Initiative Graph Format (XML-compatible)} & {2001}\; & {present} \\ 
{70} &  \href{http://en.wikipedia.org/wiki/Trivial_Graph_Format}{TGF, TGF} & \cite{trivial_graph_format} & {\checkmark } & {Trival Graph Format, and other simpleedgelists (CSV, TSV, Excel, ...)} & {NA}\; & {NA} \\ 
{71} &  \href{http://tulip.labri.fr/TulipDrupal/?q=tlp-file-format}{Tulip TLP} & \cite{auber12:_tulip_framew,tulip} & {} & {Tulip graph format} & {2002}\; & {2012} \\ 
{72} &  \href{http://www.analytictech.com/networks/dataentry.htm}{UCINET DL} & \cite{ucinet,borgatti02:_ucinet_windows} & {} & {UCINET Data Language} & {2002}\; & {2013} \\ 
{73} &  \href{http://cgi7.cs.rpi.edu/research/groups/pb/punin/public_html/XGMML/}{XGMML} & \cite{xgmml_graph_markup_model_languag} & {} & {eXtensible Graph Markup and Modeling Language} & {2000}\; & {2001} \\ 
{74} &  \href{http://www.cs.cmu.edu/~fgcozman/Research/InterchangeFormat/}{XMLBIF} & \cite{cozman13:_inter_format_bayes_networ} & {} & {XML-based BayesNets Interchange Format} & {1998}\; & {2013} \\ 
{75} &  \href{http://www.w3.org/TR/2000/NOTE-xtnd-20001121/}{XTND} & \cite{nicol00:_xtnd} & {} & {XML Transition Network Definition} & {2000}\; & {2000} \\ 
{76} &  \href{http://docs.yworks.com/yfiles/doc/developers-guide/ygf.html}{YGF} & \cite{ygf} & {\checkmark } & {Y Graph Format} & {2004}\; & {present} \\ 

%% file: excel2latex/tab2.tex
{\bf {Graph Format}} & {\bf {encoding}} & {\bf {representation}} & {\bf {structure}}  &  {\bf {integral metadata}} & {\bf {built-in compression}} \\ 

%% file: excel2latex/tab3.tex
{bintsv4} & {binary} & {edge} & {simple}  & {} & {}  \\ 
{BioGRID TAB} & {ASCII} & {edge} & {simple}  & {comments} & {}  \\ 
{BLAG, GDToolkit} & {ASCII} & {neigh} & {BNF}  & {fixed} & {}  \\ 
{BVGraph} & {binary} & {neigh} & {simple}  & {} & {\checkmark}  \\ 
{Chaco} & {ASCII} & {neigh} & {simple}  & {comments} & {}  \\ 
{Cluto} & {ASCII } & {matrix/smatrix} & {simple}  & {} & {limited}  \\ 
{DGS} & {UTF-8} & {edge} & {BNF}  & {arbitrary} & {}  \\ 
{DGML} & {Unicode} & {edge} & {XML}  & {fixed} & {}  \\ 
{DIMACS} & {ASCII} & {edge/path} & {simple}  & {fixed} & {}  \\ 
{Dot} & {UTF-8} & {edge} & {BNF}  & {arbitrary} & {}  \\ 
{DotML} & {Unicode} & {edge} & {XML}  & {arbitrary} & {}  \\ 
{DyNetML} & {Unicode} & {edge} & {XML}  & {fixed} & {}  \\ 
{GAMFF} & {ASCII} & {smatrix/matrix} & {BNF}  & {fixed} & {}  \\ 
{GDF} & {ASCII} & {edge} & {intermediate}  & {} & {}  \\ 
{GDL} & {ASCII} & {edge} & {BNF}  & {fixed} & {}  \\ 
{GEDCOM} & {UTF-8} & {neigh} & {BNF}  & {fixed} & {}  \\ 
{GEXF} & {UTF-8} & {edge} & {XML}  & {arbitrary} & {}  \\ 
{GML} & {ISO 8859} & {edge} & {BNF}  & {arbitrary} & {}  \\ 
{Graph6} & {coded ASCII} & {matrix} & {simple}  & {} & {limited}  \\ 
{Graph::Easy} & {UTF-8} & {edge/neigh} & {intermediate}  & {fixed} & {}  \\ 
{GraphEd} & {ASCII} & {neigh} & {BNF}  & {} & {}  \\ 
{GraphJSON} & {UTF-8} & {edge} & {JSON}  & {} & {}  \\ 
{GraphML} & {Unicode} & {edge} & {XML}  & {arbitrary} & {}  \\ 
{GraphSON} & {UTF-8} & {edge} & {JSON}  & {arbitrary} & {}  \\ 
{GraphXML} & {Unicode} & {edge} & {XML}  & {arbitrary} & {}  \\ 
{GraX} & {Unicode} & {edge/neigh} & {XML}  & {} & {}  \\ 
{GRXL} & {Unicode} & {edge} & {XML}  & {arbitrary} & {}  \\ 
{GT-ITM} & {ASCII} & {edge} & {simple}  & {} & {}  \\ 
{GXL} & {Unicode} & {edge} & {XML}  & {arbitrary} & {}  \\ 
{Harwell-Boeing} & {ASCII} & {smatrix} & {simple}  & {} & {}  \\ 
{Inet} & {ASCII} & {edge} & {simple}  & {} & {}  \\ 
{ITDK} & {ASCII} & {edge} & {simple}  & {comments} & {}  \\ 
{JSON Graph} & {ASCII} & {edge} & {JSON}  & {arbitrary} & {}  \\ 
{LEDA} & {ASCII} & {edge} & {simple}  & {} & {}  \\ 
{LGF} & {ASCII} & {edge} & {intermediate}  & {arbitrary} & {}  \\ 
{LGL} & {ASCII} & {neigh} & {simple}  & {} & {}  \\ 
{LibSea} & {ASCII} & {edge/path} & {BNF}  & {arbitrary} & {}  \\ 
{KrackPlot} & {ASCII} & {matrix } & {simple}  & {} & {}  \\ 
{Matlab} & {binary} & {matrix/smatrix} & {HDF5}  & {arbitrary} & {\checkmark}  \\ 
{Matrix} & {ASCII} & {smatrix } & {simple}  & {comments} & {}  \\ 
{Mivia} & {ASCII/binary} & {edge} & {simple}  & {comments} & {}  \\ 
{MultiNet} & {ASCII} & {edge/matrix} & {intermediate}  & {} & {}  \\ 
{Netdraw VNA} & {ASCII } & {edge} & {simple }  & {} & {}  \\ 
{NetML} & {Unicode} & {edge/const/proc} & {SGML}  & {fixed} & {}  \\ 
{Ncol} & {ASCII} & {edge} & {simple}  & {} & {}  \\ 
{NNF} & {ASCII} & {edge} & {simple}  & {} & {}  \\ 
{Nod} & {ASCII } & {neigh} & {simple}  & {} & {}  \\ 
{NOS} & {ASCII } & {matrix } & {simple}  & {} & {}  \\ 
{ns-tcl} & {ASCII} & {edge/procedural} & {Tcl}  & {} & {}  \\ 
{OGDL} & {ASCII (+ 8-bit var.s) and binary} & {edge/paths} & {BNF}  & {comments} & {}  \\ 
{OGML} & {Unicode} & {edge} & {XML}  & {fixed} & {}  \\ 
{Osprey} & {ASCII } & {edge} & {simple}  & {} & {}  \\ 
{Otter} & {ASCII } & {edge} & {intermediate}  & {fixed} & {}  \\ 
{Pajek (.net)} & {UTF-8} & {edge/neigh/matrix} & {intermediate}  & {comments} & {}  \\ 
{Pajek (.paj)} & {UTF-8} & {edge/neigh/matrix} & {intermediate}  & {comments } & {}  \\ 
{Planar} & {binary } & {neigh} & {simple}  & {} & {}  \\ 
{PSI MI} & {Unicode} & {edge} & {XML}  & {arbitrary} & {}  \\ 
{RSF} & {ASCII } & {edge} & {BNF}  & {comments} & {}  \\ 
{Rocketfuel} & {ASCII} & {edge/path} & {intermediate}  & {} & {}  \\ 
{Rutherford-Boeing} & {ASCII} & {smatrix } & {intermediate }  & {fixed} & {limited}  \\ 
{SGB} & {ASCII} & {edge/neigh} & {intermediate}  & {fixed} & {}  \\ 
{SGF} & {Unicode} & {edge} & {XML}  & {arbitrary} & {}  \\ 
{S-Dot} & {ASCII} & {edge/procedural} & {lisp}  & {arbitrary} & {}  \\ 
{SIF} & {ASCII (URL encode)} & {edge/neigh} & {simple}  & {} & {}  \\ 
{SNAP} & {ASCII} & {edge} & {simple}  & {comments} & {}  \\ 
{SoNIA} & {ASCII} & {edge} & {intermediate}  & {comments} & {}  \\ 
{Sparse6} & {coded ASCII} & {neigh} & {simple}  & {} & {limited}  \\ 
{StOCNET} & {ASCII} & {matrix} & {simple}  & {} & {}  \\ 
{TEI} & {Unicode} & {edge} & {XML}  & {fixed} & {}  \\ 
{TGF, TGF} & {ASCII} & {edge} & {simple}  & {comments} & {}  \\ 
{Tulip TLP} & {ASCII} & {edge} & {BNF}  & {fixed} & {}  \\ 
{UCINET DL} & {ASCII} & {neigh/edge/matrix} & {intermediate}  & {} & {}  \\ 
{XGMML} & {Unicode} & {edge} & {XML}  & {arbitrary} & {}  \\ 
{XMLBIF} & {Unicode } & {neigh} & {XML}  & {arbitrary} & {}  \\ 
{XTND} & {Unicode} & {edge} & {XML}  & {comments} & {}  \\ 
{YGF} & {binary} & {edge} & {simple}  & {} & {\checkmark}  \\ 

%% file: excel2latex/tab4.tex
{\bf {Graph Format}} & {\bf {directed}} & {\bf {multi-graph}} & {\bf {hyper-graph}} & {\bf {hierarchy}} & {\bf {meta-graph}} & {\bf {edge-edge}}\\ 

%% file: excel2latex/tab5.tex
{bintsv4} & {directed} & {} & {} & {} & {} & {} \\ 
{BioGRID TAB} & {directed} & {} & {} & {} & {} & {} \\ 
{BLAG, GDToolkit} & {either} & {} & {} & {\checkmark} & {} & {} \\ 
{BVGraph} & {directed} & {} & {} & {} & {} & {} \\ 
{Chaco} & {undirected} & {} & {} & {} & {} & {} \\ 
{Cluto} & {directed} & {} & {} & {} & {} & {} \\ 
{DGS} & {mixed} & {\checkmark} & {} & {} & {} & {} \\ 
{DGML} & {unspecified} & {unspecified} & {} & {} & {} & {} \\ 
{DIMACS} & {either} & {\checkmark} & {} & {} & {} & {} \\ 
{Dot} & {mixed} & {\checkmark} & {\checkmark} & {\checkmark} & {} & {} \\ 
{DotML} & {mixed} & {\checkmark} & {\checkmark} & {\checkmark} & {} & {} \\ 
{DyNetML} & {directed} & {unspecified} & {} & {} & {} & {} \\ 
{GAMFF} & {either} & {unspecified} & {\checkmark} & {} & {} & {} \\ 
{GDF} & {unspecified} & {unspecified} & {} & {} & {} & {} \\ 
{GDL} & {directed } & {unspecified} & {} & {} & {} & {} \\ 
{GEDCOM} & {mixed} & {unspecified} & {} & {} & {} & {} \\ 
{GEXF} & {mixed} & {\checkmark} & {} & {\checkmark} & {} & {} \\ 
{GML} & {either} & {\checkmark} & {} & {} & {} & {} \\ 
{Graph6} & {directed} & {loops only} & {} & {} & {} & {} \\ 
{Graph::Easy} & {mixed} & {\checkmark} & {\checkmark} & {\checkmark} & {} & {} \\ 
{GraphEd} & {unspecified} & {\checkmark} & {} & {} & {} & {} \\ 
{GraphJSON} & {mixed} & {unspecified} & {} & {} & {} & {} \\ 
{GraphML} & {mixed} & {\checkmark} & {\checkmark} & {\checkmark} & {} & {\checkmark} \\ 
{GraphSON} & {directed} & {unspecified} & {} & {} & {} & {} \\ 
{GraphXML} & {either} & {unspecified} & {} & {\checkmark} & {} & {} \\ 
{GraX} & {directed} & {unspecified} & {} & {} & {} & {} \\ 
{GRXL} & {directed} & {unspecified} & {} & {} & {} & {} \\ 
{GT-ITM} & {undirected} & {} & {} & {} & {} & {} \\ 
{GXL} & {mixed} & {\checkmark} & {\checkmark} & {\checkmark} & {} & {} \\ 
{Harwell-Boeing} & {directed} & {} & {} & {} & {} & {} \\ 
{Inet} & {undirected} & {} & {} & {} & {} & {} \\ 
{ITDK} & {undirected} & {} & {} & {} & {} & {} \\ 
{JSON Graph} & {either} & {\checkmark} & {} & {} & {} & {} \\ 
{LEDA} & {either} & {unspecified} & {} & {} & {} & {} \\ 
{LGF} & {either} & {unspecified} & {} & {} & {} & {} \\ 
{LGL} & {undirected} & {unspecified} & {} & {} & {} & {} \\ 
{LibSea} & {directed } & {unspecified} & {} & {} & {} & {} \\ 
{KrackPlot} & {directed } & {} & {} & {} & {} & {} \\ 
{Matlab} & {directed} & {} & {} & {} & {} & {} \\ 
{Matrix} & {either/bipartite} & {} & {} & {} & {} & {} \\ 
{Mivia} & {directed} & {unspecified} & {} & {} & {} & {} \\ 
{MultiNet} & {unspecified} & {unspecified} & {} & {} & {} & {} \\ 
{Netdraw VNA} & {directed} & {unspecified} & {} & {} & {} & {} \\ 
{NetML} & {either} & {unspecified} & {} & {\checkmark} & {} & {} \\ 
{Ncol} & {undirected} & {} & {} & {} & {} & {} \\ 
{NNF} & {either} & {} & {} & {\checkmark} & {} & {} \\ 
{Nod} & {directed } & {} & {} & {} & {} & {} \\ 
{NOS} & {directed } & {} & {} & {} & {} & {} \\ 
{ns-tcl} & {directed} & {\checkmark} & {\checkmark} & {\checkmark} & {} & {} \\ 
{OGDL} & {directed} & {} & {} & {} & {} & {} \\ 
{OGML} & {directed} & {unspecified} & {\checkmark} & {} & {} & {} \\ 
{Osprey} & {undirected} & {} & {} & {} & {} & {} \\ 
{Otter} & {mixed} & {unspecified} & {} & {} & {} & {} \\ 
{Pajek (.net)} & {mixed} & {loops only} & {} & {} & {} & {} \\ 
{Pajek (.paj)} & {mixed} & {loops only} & {} & {\checkmark} & {} & {} \\ 
{Planar} & {planar} & {} & {} & {} & {} & {} \\ 
{PSI MI} & {unspecified} & {} & {\checkmark} & {} & {} & {} \\ 
{RSF} & {directed } & {unspecified} & {} & {} & {} & {} \\ 
{Rocketfuel} & {undirected} & {} & {} & {\checkmark} & {} & {} \\ 
{Rutherford-Boeing} & {either} & {} & {} & {} & {} & {} \\ 
{SGB} & {unspecified} & {unspecified} & {} & {} & {} & {} \\ 
{SGF} & {directed} & {unspecified} & {} & {\checkmark} & {} & {} \\ 
{S-Dot} & {mixed} & {\checkmark} & {\checkmark} & {\checkmark} & {} & {} \\ 
{SIF} & {mixed} & {} & {} & {} & {} & {} \\ 
{SNAP} & {either} & {\checkmark} & {} & {} & {} & {} \\ 
{SoNIA} & {directed} & {\checkmark} & {} & {\checkmark} & {} & {} \\ 
{Sparse6} & {undirected} & {\checkmark} & {} & {} & {} & {} \\ 
{StOCNET} & {directed} & {} & {} & {} & {} & {} \\ 
{TEI} & {either or tree} & {unspecified} & {} & {} & {} & {} \\ 
{TGF, TGF} & {either} & {} & {} & {} & {} & {} \\ 
{Tulip TLP} & {directed} & {unspecified} & {} & {\checkmark} & {} & {} \\ 
{UCINET DL} & {directed} & {} & {} & {} & {} & {} \\ 
{XGMML} & {either} & {unspecified} & {} & {} & {} & {} \\ 
{XMLBIF} & {DAG} & {} & {} & {} & {} & {} \\ 
{XTND} & {directed} & {unspecified} & {} & {} & {} & {} \\ 
{YGF} & {mixed} & {} & {} & {\checkmark} & {} & {} \\ 

%% file: excel2latex/tab6.tex
{\bf {Graph Format}} & {\bf {edgeweights}} & {\bf {multiple attributes}} & {\bf {default values}} & {\bf {multiple inheritance}} & {\bf {visualisation data}} & {\bf {ports}} & {\bf {temporal data/dynamics}} \\ 

%% file: excel2latex/tab7.tex
{bintsv4} & {} & {} & {} & {} & {} & {} & {}  \\ 
{BioGRID TAB} & {} & {fixed} & {} & {} & {} & {} & {}  \\ 
{BLAG, GDToolkit} & {} & {} & {} & {} & {\checkmark} & {} & {}  \\ 
{BVGraph} & {} & {} & {} & {} & {} & {} & {}  \\ 
{Chaco} & {\checkmark} & {} & {} & {} & {\checkmark} & {} & {}  \\ 
{Cluto} & {\checkmark} & {} & {} & {} & {} & {} & {}  \\ 
{DGS} & {\checkmark} & {arbitrary} & {} & {} & {\checkmark} & {} & {\checkmark}  \\ 
{DGML} & {\checkmark} & {arbitrary} & {\checkmark} & {\checkmark} & {\checkmark} & {} & {}  \\ 
{DIMACS} & {\checkmark} & {} & {} & {} & {\checkmark} & {} & {}  \\ 
{Dot} & {\checkmark} & {arbitrary} & {\checkmark} & {} & {\checkmark} & {\checkmark} & {}  \\ 
{DotML} & {\checkmark} & {arbitrary} & {\checkmark} & {} & {\checkmark} & {\checkmark} & {}  \\ 
{DyNetML} & {\checkmark} & {arbitrary} & {} & {} & {} & {} & {\checkmark}  \\ 
{GAMFF} & {\checkmark} & {fixed} & {} & {} & {\checkmark} & {} & {}  \\ 
{GDF} & {\checkmark} & {arbitrary} & {} & {} & {\checkmark} & {} & {}  \\ 
{GDL} & {\checkmark} & {fixed} & {\checkmark} & {} & {\checkmark} & {} & {}  \\ 
{GEDCOM} & {} & {fixed} & {} & {} & {} & {} & {}  \\ 
{GEXF} & {\checkmark} & {arbitrary} & {\checkmark} & {\checkmark} & {\checkmark} & {} & {\checkmark}  \\ 
{GML} & {\checkmark} & {arbitrary} & {} & {} & {\checkmark} & {} & {}  \\ 
{Graph6} & {} & {} & {} & {} & {} & {} & {}  \\ 
{Graph::Easy} & {\checkmark} & {fixed} & {\checkmark} & {} & {\checkmark} & {\checkmark} & {}  \\ 
{GraphEd} & {} & {} & {} & {} & {\checkmark} & {} & {}  \\ 
{GraphJSON} & {\checkmark} & {arbitrary} & {\checkmark} & {\checkmark} & {\checkmark} & {} & {}  \\ 
{GraphML} & {\checkmark} & {arbitrary} & {\checkmark} & {visualisation} & {\checkmark} & {\checkmark} & {}  \\ 
{GraphSON} & {\checkmark} & {arbitrary} & {} & {} & {} & {} & {}  \\ 
{GraphXML} & {\checkmark} & {arbitrary} & {} & {} & {\checkmark} & {} & {\checkmark}  \\ 
{GraX} & {\checkmark} & {arbitrary} & {} & {} & {} & {} & {}  \\ 
{GRXL} & {\checkmark} & {arbitrary} & {\checkmark} & {} & {\checkmark} & {} & {}  \\ 
{GT-ITM} & {\checkmark} & {fixed} & {} & {} & {\checkmark} & {} & {}  \\ 
{GXL} & {\checkmark} & {arbitrary} & {} & {} & {} & {} & {}  \\ 
{Harwell-Boeing} & {\checkmark} & {} & {} & {} & {} & {} & {}  \\ 
{Inet} & {\checkmark} & {} & {} & {} & {\checkmark} & {} & {}  \\ 
{ITDK} & {} & {} & {} & {} & {} & {} & {}  \\ 
{JSON Graph} & {\checkmark} & {arbitrary} & {} & {} & {} & {} & {}  \\ 
{LEDA} & {\checkmark} & {fixed} & {} & {} & {} & {} & {}  \\ 
{LGF} & {\checkmark} & {\checkmark} & {} & {} & {} & {} & {}  \\ 
{LGL} & {\checkmark} & {} & {} & {} & {} & {} & {}  \\ 
{LibSea} & {\checkmark} & {\checkmark} & {\checkmark} & {} & {\checkmark} & {} & {}  \\ 
{KrackPlot} & {} & {} & {} & {} & {\checkmark} & {} & {}  \\ 
{Matlab} & {\checkmark} & {almost} & {} & {} & {} & {} & {}  \\ 
{Matrix} & {\checkmark} & {} & {} & {} & {} & {} & {}  \\ 
{Mivia} & {\checkmark} & {} & {} & {} & {} & {} & {}  \\ 
{MultiNet} & {\checkmark} & {arbitrary} & {} & {} & {} & {} & {}  \\ 
{Netdraw VNA} & {\checkmark} & {arbitrary} & {} & {} & {} & {} & {}  \\ 
{NetML} & {\checkmark} & {\checkmark} & {\checkmark} & {} & {\checkmark} & {} & {}  \\ 
{Ncol} & {\checkmark} & {} & {} & {} & {} & {} & {}  \\ 
{NNF} & {} & {} & {} & {} & {} & {} & {}  \\ 
{Nod} & {} & {} & {} & {} & {} & {} & {}  \\ 
{NOS} & {\checkmark} & {} & {} & {} & {} & {} & {}  \\ 
{ns-tcl} & {\checkmark} & {fixed} & {} & {} & {\checkmark} & {\checkmark} & {\checkmark}  \\ 
{OGDL} & {} & {} & {} & {} & {} & {} & {}  \\ 
{OGML} & {?} & {fixed} & {} & {} & {\checkmark} & {\checkmark} & {}  \\ 
{Osprey} & {} & {fixed} & {} & {} & {} & {} & {}  \\ 
{Otter} & {} & {arbitrary} & {} & {} & {\checkmark} & {} & {}  \\ 
{Pajek (.net)} & {\checkmark} & {fixed} & {visualisation} & {} & {\checkmark} & {\checkmark} & {\checkmark}  \\ 
{Pajek (.paj)} & {\checkmark} & {arbitrary} & {visualisation} & {} & {\checkmark} & {\checkmark} & {\checkmark}  \\ 
{Planar} & {} & {} & {} & {} & {} & {} & {}  \\ 
{PSI MI} & {\checkmark} & {arbitrary} & {} & {} & {} & {} & {}  \\ 
{RSF} & {\checkmark} & {arbitrary} & {} & {} & {} & {} & {}  \\ 
{Rocketfuel} & {\checkmark} & {} & {} & {} & {\checkmark} & {} & {}  \\ 
{Rutherford-Boeing} & {\checkmark} & {fixed} & {} & {} & {} & {} & {}  \\ 
{SGB} & {\checkmark} & {fixed} & {} & {} & {} & {} & {}  \\ 
{SGF} & {\checkmark} & {arbitrary} & {} & {} & {} & {} & {}  \\ 
{S-Dot} & {\checkmark} & {arbitrary} & {\checkmark} & {} & {\checkmark} & {\checkmark} & {}  \\ 
{SIF} & {} & {} & {} & {} & {} & {} & {}  \\ 
{SNAP} & {} & {} & {} & {} & {} & {} & {}  \\ 
{SoNIA} & {\checkmark} & {arbitrary} & {} & {} & {\checkmark} & {} & {\checkmark}  \\ 
{Sparse6} & {} & {} & {} & {} & {} & {} & {}  \\ 
{StOCNET} & {\checkmark} & {fixed} & {} & {} & {} & {} & {}  \\ 
{TEI} & {\checkmark} & {fixed} & {} & {} & {} & {} & {}  \\ 
{TGF, TGF} & {} & {} & {} & {} & {} & {} & {}  \\ 
{Tulip TLP} & {\checkmark} & {arbitrary} & {\checkmark} & {} & {} & {} & {}  \\ 
{UCINET DL} & {\checkmark} & {} & {} & {} & {} & {} & {}  \\ 
{XGMML} & {\checkmark} & {arbitrary} & {} & {} & {} & {} & {}  \\ 
{XMLBIF} & {\checkmark} & {fixed} & {} & {} & {} & {} & {}  \\ 
{XTND} & {} & {fixed} & {} & {} & {} & {} & {}  \\ 
{YGF} & {\checkmark} & {arbitrary} & {} & {\checkmark} & {\checkmark} & {} & {}  \\ 

%% file: excel2latex/tab8.tex
{\bf {Graph Format}} &  {\bf {extensible}} & {\bf {schema checking}} & {\bf {checksums}} & {\bf {external data references}}  & {\bf {multiple graphs}} & {\bf {incremental specifications}} \\ 

%% file: excel2latex/tab9.tex
{bintsv4} & {} & {} & {} & {} & {} & {} \\ 
{BioGRID TAB} & {} & {} & {} & {\checkmark} & {} & {} \\ 
{BLAG, GDToolkit} & {} & {} & {} & {} & {} & {} \\ 
{BVGraph} & {} & {} & {} & {} & {} & {} \\ 
{Chaco} & {} & {} & {} & {} & {} & {} \\ 
{Cluto} & {} & {} & {} & {} & {} & {} \\ 
{DGS} & {} & {} & {} & {CSS} & {} & {} \\ 
{DGML} & {} & {\checkmark} & {} & {} & {} & {} \\ 
{DIMACS} & {} & {} & {\checkmark} & {} & {\checkmark} & {} \\ 
{Dot} & {} & {\checkmark} & {} & {} & {} & {} \\ 
{DotML} & {} & {} & {} & {} & {} & {} \\ 
{DyNetML} & {} & {\checkmark} & {} & {} & {\checkmark} & {\checkmark} \\ 
{GAMFF} & {} & {} & {} & {} & {} & {} \\ 
{GDF} & {\checkmark} & {} & {} & {} & {} & {} \\ 
{GDL} & {} & {} & {} & {} & {} & {} \\ 
{GEDCOM} & {\checkmark} & {} & {} & {\checkmark} & {} & {} \\ 
{GEXF} & {\checkmark} & {\checkmark} & {} & {\checkmark} & {\checkmark} & {} \\ 
{GML} & {} & {} & {} & {} & {} & {} \\ 
{Graph6} & {} & {} & {} & {} & {\checkmark} & {\checkmark} \\ 
{Graph::Easy} & {} & {} & {} & {\checkmark} & {} & {} \\ 
{GraphEd} & {} & {} & {} & {} & {} & {} \\ 
{GraphJSON} & {} & {\checkmark} & {} & {} & {} & {} \\ 
{GraphML} & {\checkmark} & {\checkmark} & {} & {\checkmark} & {\checkmark} & {} \\ 
{GraphSON} & {\checkmark} & {\checkmark} & {} & {} & {} & {} \\ 
{GraphXML} & {\checkmark} & {\checkmark} & {} & {\checkmark} & {\checkmark} & {} \\ 
{GraX} & {} & {\checkmark} & {} & {} & {} & {} \\ 
{GRXL} & {\checkmark} & {\checkmark} & {} & {} & {} & {} \\ 
{GT-ITM} & {} & {} & {} & {} & {} & {} \\ 
{GXL} & {\checkmark} & {\checkmark} & {} & {\checkmark} & {\checkmark} & {} \\ 
{Harwell-Boeing} & {} & {} & {} & {} & {} & {} \\ 
{Inet} & {} & {} & {} & {} & {} & {} \\ 
{ITDK} & {} & {} & {} & {} & {} & {} \\ 
{JSON Graph} & {} & {\checkmark} & {} & {} & {\checkmark} & {} \\ 
{LEDA} & {} & {} & {} & {} & {} & {} \\ 
{LGF} & {} & {} & {} & {} & {} & {} \\ 
{LGL} & {} & {} & {} & {} & {} & {} \\ 
{LibSea} & {} & {\checkmark} & {} & {} & {?} & {} \\ 
{KrackPlot} & {} & {} & {} & {} & {} & {} \\ 
{Matlab} & {} & {} & {} & {} & {} & {} \\ 
{Matrix} & {} & {} & {} & {} & {} & {} \\ 
{Mivia} & {} & {} & {} & {} & {} & {} \\ 
{MultiNet} & {} & {} & {} & {} & {} & {} \\ 
{Netdraw VNA} & {} & {} & {} & {} & {} & {} \\ 
{NetML} & {} & {\checkmark} & {} & {} & {\checkmark} & {\checkmark} \\ 
{Ncol} & {} & {} & {} & {} & {} & {} \\ 
{NNF} & {} & {} & {} & {} & {} & {} \\ 
{Nod} & {} & {} & {} & {} & {} & {} \\ 
{NOS} & {} & {} & {} & {} & {\checkmark} & {} \\ 
{ns-tcl} & {\checkmark} & {} & {} & {} & {} & {} \\ 
{OGDL} & {} & {\checkmark} & {} & {} & {} & {} \\ 
{OGML} & {} & {} & {} & {} & {} & {} \\ 
{Osprey} & {} & {} & {} & {} & {} & {} \\ 
{Otter} & {} & {} & {} & {} & {} & {} \\ 
{Pajek (.net)} & {\checkmark} & {} & {} & {} & {} & {} \\ 
{Pajek (.paj)} & {\checkmark} & {} & {} & {} & {} & {} \\ 
{Planar} & {} & {} & {} & {} & {} & {} \\ 
{PSI MI} & {\checkmark} & {\checkmark} & {} & {\checkmark} & {} & {} \\ 
{RSF} & {} & {} & {} & {} & {} & {} \\ 
{Rocketfuel} & {} & {} & {} & {} & {\checkmark} & {} \\ 
{Rutherford-Boeing} & {} & {} & {\checkmark} & {} & {} & {} \\ 
{SGB} & {} & {partial} & {\checkmark} & {} & {} & {} \\ 
{SGF} & {} & {\checkmark} & {} & {} & {} & {} \\ 
{S-Dot} & {} & {\checkmark} & {} & {} & {} & {} \\ 
{SIF} & {} & {} & {} & {} & {} & {} \\ 
{SNAP} & {} & {} & {} & {} & {} & {} \\ 
{SoNIA} & {} & {} & {} & {} & {} & {} \\ 
{Sparse6} & {} & {} & {} & {} & {\checkmark} & {\checkmark} \\ 
{StOCNET} & {} & {} & {} & {} & {} & {} \\ 
{TEI} & {} & {\checkmark} & {} & {} & {} & {} \\ 
{TGF, TGF} & {} & {} & {} & {} & {} & {} \\ 
{Tulip TLP} & {} & {} & {} & {} & {} & {} \\ 
{UCINET DL} & {} & {} & {} & {} & {} & {} \\ 
{XGMML} & {} & {\checkmark} & {} & {} & {} & {} \\ 
{XMLBIF} & {} & {\checkmark} & {} & {} & {} & {} \\ 
{XTND} & {} & {} & {} & {} & {} & {} \\ 
{YGF} & {\checkmark} & {} & {} & {} & {} & {} \\ 

%% file: excel2latex/tab10.tex
{\bf {Dataset}} & & {\bf {Full name}} & {\bf {Format}} \\ 

%% file: excel2latex/tab11.tex
\href{http://mivia.unisa.it/datasets/graph-database/}{ARG/VF} & \cite{mivia_graph_datab} & {Mivia ARG Database and VF Library} & {Mivia} \\ 
\href{http://thebiogrid.org/}{BioGRID} & \cite{chatr-aryamontri12:_biogr_inter_datab} & {Biological General Repository for Interaction Datasets} & {PSI MI, Osprey, BioGRID, PTMTAB} \\ 
\href{http://www.casos.cs.cmu.edu/computational_tools/data2.php}{CASOS} & \cite{cmu_casos} & {CMU CASOS Datasets} & {DyNetML, GML, UCINET, GraphML} \\ 
\href{http://boston.lti.cs.cmu.edu/clueweb09/wiki/tiki-index.php?page=Web+Graph}{ClueWeb09} & \cite{clueweb} & {ClueWeb09 Web Graph} & {BVGraph, TGF} \\ 
\href{http://dimacs.rutgers.edu/Challenges/}{DIMACS10} & \cite{dimac_implem_chall} & {DIMACS Implementation Challenges} & {DIMACS} \\ 
\href{http://law.di.unimi.it/datasets.php}{DSI} & \cite{BCSU3,labor_web_algor} & {Web Algorithmics Lab Data} & {BVGraph } \\ 
\href{http://cis.jhu.edu/~parky/Enron/}{Enron} & \cite{priebe11:_scan_statis_enron_graph,park10:_scan_statis_enron_graph} & {Enron email dataset} & {TGF} \\ 
\href{http://www.graphdrawing.org/data.html}{Graph-Archive} & \cite{graphdrawing} & {GraphArchive - Exchange and Archive System for Graphs} & {GraphML} \\ 
\href{https://github.com/uwsampa/graphbench/wiki}{GraphBench} & \cite{graphbench} & {GraphBench} & {TGF} \\ 
\href{http://hog.grinvin.org/}{HOG} & \cite{hog} & {The House of Graphs} & {TGF, Graph6, Multicode, Planar} \\ 
\href{http://www.hprd.org/FAQ/index_html, http://www.ncbi.nlm.nih.gov/pmc/articles/PMC1347503/}{HPRD} & \cite{mishra06:_human} & {Human Protein Reference Database} & {PSI MI, TSV } \\ 
\href{http://webdatacommons.org/hyperlinkgraph/}{Hyperlink} & \cite{meusel14:_graph_struc_web_revis} & {Web Data Commons - Hyperlink Graphs} & {Pajek, WebGraph} \\ 
\href{http://www.iam.unibe.ch/fki/databases/iam-graph-database}{IAM} & \cite{riesen08:_iam} & {IAM Graph Database Repository } & {GXL} \\ 
\href{http://www.caida.org/data/internet-topology-data-kit/}{ITDK} & \cite{14:itdk} & {CAIDA Macroscopic Internet Topology Data Kit} & {ITDK} \\ 
\href{http://www.topology-zoo.org/index.html}{Zoo} & \cite{Zoo} & {Internet Topology Zoo} & {GML, GraphML} \\ 
\href{http://math.nist.gov/MatrixMarket/}{Matrix Market} & \cite{boisvert96:_matrix_market} & {Matrix Market} & {Matrix Market} \\ 
\href{http://www.nas.nasa.gov/Software/GAMFF/}{NAS} & \cite{nas_graph_collec} & {NAS (NASA) Graph Collection} & {GAMFF} \\ 
\href{http://pajek.imfm.si/doku.php?id=data:pajek:index}{Pajek} & \cite{13:_pajek} & {Pajek Data Sets} & {Pajek} \\ 
\href{http://www.cs.washington.edu/research/projects/networking/www/rocketfuel/}{Rocketfuel} & \cite{rocketfuel_0} & {Rocketfuel} & {Rocketfuel} \\ 
\href{http://people.sc.fsu.edu/~jburkardt/datasets/sgb/sgb.html}{SGB} & \cite{08:_datas_stanf_graph_base,knuth:_stanf_graph} & {Stanford GraphBase} & {SGB} \\ 
\href{http://snap.stanford.edu/snap/}{SNAP} & \cite{snap} & {Stanford Network Analysis Platform} & {SNAP} \\ 
\href{http://toreopsahl.com/datasets/}{Tore} & \cite{opsahl:_dataset} & {Tore Opsahl Datsets} & {UCINET, tnet} \\ 
\href{http://an.kaist.ac.kr/traces/WWW2010.html}{Twitter} & \cite{kwak10:_twitter} & {What is Twitter, a Social Network or a News Media?} & {TGF} \\ 
\href{http://www.cise.ufl.edu/research/sparse/matrices/}{UF} & \cite{davis11:_florida} & {The University of Florida Sparse Matrix Collection } & {Matrix Market, Rutherford-Boeing, Matlab} \\ 
\href{http://vlado.fmf.uni-lj.si/pub/networks/data/WaFa/default.htm}{WF} & \cite{wasserman94:_social_networ_analy} & {Wasserman and Faust datasets} & {Pajek} \\ 